\documentclass[10pt,twocolumn]{IEEEtran}

\usepackage{amsmath,amssymb,amsfonts,amsthm}
\usepackage{algorithmic}
\usepackage{graphicx}
\usepackage[caption=false]{subfig}
\usepackage{algorithm,algorithmic}
\usepackage{hyperref}
\usepackage{textcomp}

\newcommand{\tp}{\normalfont \text{T}}

\newcommand{\Val}{\operatorname{Val}}
\newtheorem{definition}{Definition}

\newtheorem{remark}{Remark}
\newtheorem{lemma}{Lemma}
\newtheorem{cor}{Corollary}
\newtheorem{proposition}{Proposition}
\newtheorem{theorem}{Theorem}
\newtheorem{assumption}{Assumption}

\def\qedsymbol{\ensuremath{\Box}}      

\def\qed{\ifhmode\unskip\nobreak\fi\quad\qedsymbol}     
\def\frqed{\ifhmode\nobreak\hbox to5pt{\hfil}\nobreak%
	\hskip 0pt plus1fill\nobreak\fi\quad\qedsymbol\renewcommand{\qed}{}} 

\def\QEDsymbol{\vrule width.6em height.5em depth.1em\relax}
\def\frQED{\ifhmode\nobreak\hbox to5pt{\hfil}\nobreak%
	\hskip 0pt plus1fill\nobreak\fi\quad\QEDsymbol\renewcommand{\qed}{}} 
\def\QED{\ifhmode\unskip\nobreak\fi\quad\QEDsymbol}     

\begin{document}
\title{\texttt{FlipDyn} with Control: Resource Takeover Games with Dynamics}
\author{Sandeep Banik, \IEEEmembership{Member, IEEE} and Shaunak D. Bopardikar, \IEEEmembership{Senior Member, IEEE}
\thanks{This research was supported in part by the NSF Award CNS-2134076 under the Secure and Trustworthy Cyberspace (SaTC) program and in part by the NSF CAREER Award ECCS-2236537.}
\thanks{The first author is with the Department of Mechanical Engineering at University of Illinois Urbana-Champaign, IL, USA. Email: \texttt{baniksan@illinois.edu}.}
\thanks{The second author is with the Department of Electrical and Computer Engineering at Michigan State University, East Lansing, MI, USA. Email: \texttt{shaunak@egr.msu.edu}.}
}

\maketitle


\begin{abstract}
We introduce \texttt{FlipDyn} with control, a finite-horizon zero-sum resource takeover game, where a defender and an adversary decide when to takeover and how to control a common resource. 
At each discrete-time step, the players can take over or retain control, incurring state and control-dependent costs. 
The system is modeled as a hybrid dynamical system, with a discrete \texttt{FlipDyn} state determining control authority. 
Our contributions are: (i) For arbitrary non-negative costs, we derive the saddle-point value of the \texttt{FlipDyn} game and the corresponding Nash equilibria (NE) takeover strategies. 
(ii) For linear dynamical systems with quadratic costs, we establish sufficient conditions under which the game admits an NE. 
(iii) For scalar linear dynamical systems with quadratic costs, we derive parameterized NE takeover strategies and saddle-point values independent of the continuous state. 
(iv) For higher-dimensional linear dynamical systems with quadratic costs, we derive approximate NE takeover strategies and control policies, and compute bounds on the saddle-point values. 
We validate our results through a numerical study on adversarial control of a linear system. 
\end{abstract}


\section{Introduction}
\label{sec:introduction}
\IEEEPARstart{T}{he} integration of cyber and physical systems, driven by advancements in automation, computation, and communication technologies has transformed numerous industries, such as medical devices, traffic control, industrial systems, power grids, and autonomous vehicles~\cite{rajkumar2010cyber,baheti2011cyber,lee2016introduction}.
However, this connectivity has also amplified adversarial risks, with malicious actors exploiting system vulnerabilities~\cite{cardenas2008research,parkinson2017cyber,lun2019state}.
To mitigate these risks, new approaches combining game theory~\cite{fotiadis2022concurrent,tushar2023survey,zhu2015game}, control theory~\cite{bai2017kalman,lun2019state}, and machine learning~\cite{wickramasinghe2018generalization} have emerged to design resilient defense strategies.

The security attributes of any cyber-physical system (CPS) are broadly classified into three categories: confidentiality, integrity, and availability~\cite{avizienis2004basic}. 
In this work, we consider an adversary who targets both confidentiality and integrity by taking control of a dynamical system when it enters a vulnerable state. 
The adversary then sends malicious control signals~\cite{liu2021flipit} to drive the system to undesirable states~\cite{kontouras2014adversary,kontouras2015covert}. Such actions can cause permanent damage, disrupt services, and lead to operational losses.
Therefore, it is imperative to develop defensive strategies that continuously detect and counter adversarial behavior while balancing operating costs and system performance. 
This paper introduces a framework that models the problem of dynamic resource takeovers and designs defense policies with guarantees on their performance.

The framework of \texttt{FlipIT}~\cite{van2013flipit}, a game of resource takeovers, was introduced to model a conflict between a defender and an adversary competing over a common resource, such as a computing device or cloud service~\cite{bowers2012defending}. 
This framework was extended to incorporate dynamic environments with varying costs and attack success probabilities~\cite{johnson2015games}. 
\texttt{FlipIT} was then generalized to multiple resources, referred to as \texttt{FlipThem}~\cite{laszka2014flipthem}, along with a variation that allows the defender to configure resources to deter adversarial attacks beyond a certain threshold~\cite{leslie2015threshold}. 
Resource constraints were added~\cite{zhang2020defending} as part of a two-player non-zero-sum game for multiple resource takeovers, along with a threshold-based takeover model for critical infrastructure systems~\cite{canzani2016cyber}.
\texttt{FlipIT} was extended to graphs, termed \texttt{FlipNet}~\cite{saha2017flipnet}, to explore graph structures, best-response strategies, and Nash equilibria.  
Beyond cybersecurity, \texttt{FlipIT} was applied to supervisory control and data acquisition (SCADA) systems~\cite{liu2021flipit} to assess the impact of cyberattacks involving insider assistance.
In addition to these developments, \texttt{FlipIT} has been applied broadly across system security to address diverse threats and defense strategies~\cite{bowers2012defending}.
Notably, the aforementioned works primarily focused on resource takeovers of a \emph{static system}, ignoring the dynamic evolution of physical systems.
In contrast, our work incorporates the \emph{dynamics} of a physical system in the game of resource takeovers between an adversary and a defender.

The framework in~\cite{ding2013stochastic} addresses the synthesis of safety controls in stochastic hybrid systems over a finite-horizon, as a stochastic game. 
Our work considers a discrete-time game with two hybrid states, but with a key distinction: only one player controls the system in a given hybrid state, while allowing for a potential switch to the other state.
A related investigation into safe controller design within two hybrid states was conducted in~\cite{dallal2016synthesis}, modeling a game between a controller aiming to enforce safety and an environment attempting to violate it.
In~\cite{fiscko2019control}, a multi-player game was introduced, where a superplayer manages a parameterized utility of all players to derive cost-optimal policies. 
 Similarly,~\cite{fiscko2021efficient} studied multi-agent systems clustered under a superplayer to synthesize a cluster-based control policy.
The aforementioned works correspond to the special case of two clusters in our setting without any \emph{coupling}. In contrast, our work addresses control policy design in the presence of coupling between clusters.

The works in~\cite{ivanov2023fast, vrabie2010adaptive, bacsar2017riccati, luo2020policy} formulate two-player zero-sum games for linear dynamical systems as Riccati  equations in continuous time and discrete-time.
Analytical and offline solutions for such games with known dynamics were presented for finite-horizon~\cite{bacsar1998dynamic} and for infinite-horizon quadratic costs~\cite{bacsar2017riccati, ivanov2023fast}.
To address unknown dynamics, adaptive dynamic programming~\cite{vrabie2010adaptive} and Q-learning~\cite{luo2020policy} were introduced.
Extensions to infinite-horizon nonlinear dynamics with quadratic costs were proposed in~\cite{ren2022zero, zhong2017model}, while switching dynamics in zero-sum games were explored in~\cite{lv2020two, wang2024optimal}. 
Compared to the aforementioned models, our paper simultaneously solves for both coupled value functions and control policies for both players, incorporating discrete takeover actions into the zero-sum game framework.

The setup in~\cite{kontouras2014adversary} closely resembles our \texttt{FlipDyn}~\cite{FlipDyn_banik2022} framework, but is greatly limited in assuming periodic policies with only scalar inputs. 
Building on this,\cite{kontouras2015covert} considers multi-dimensional controls and designs contractive policies against covert attacks under state and input constraints.
Related work explores covert misappropriation via feedback~\cite{smith2015covert} and covert attacks on load frequency control systems using reference signals~\cite{mohan2020covert}. 
Our approach provides a feedback mechanism to infer control authority and enables takeover at any instant, balancing cost and performance in a game-theoretic setting.

Recent studies have shown that adversaries can intermittently take control of CPS, altering their dynamics.
Denial-of-Service (DoS) attacks on remotely controlled LTI systems~\cite{shishehforoushEventtriggeredControlLinear2012} and input-to-state stability under DoS~\cite{depersisInputtoStateStabilizingControl2015} align with our framework, where takeovers resemble jamming events. 
These examples underscore the need to model and mitigate dynamic takeovers, especially as autonomous systems become more integrated into modern infrastructure.

Our prior works~\cite{FlipDyn_banik2022} and~\cite{FlipDynG_2025} introduced the game of resource takeover in dynamical systems, with known control policies, and in graph-based setup with multiple \texttt{FlipDyn} states. In this paper, we extend this framework by simultaneously computing both the takeover strategies and control policies for each player. The main contributions are as follows:

\begin{enumerate}
    \item \textbf{Takeover strategies for any discrete-time dynamical system}: We formulate a two-player zero-sum takeover game between a defender and an adversary seeking to control a discrete-time dynamical system. 
    This game encompasses dynamic takeovers, with state and control-based costs. Under the assumption of a prior known control policies over the finite-horizon, we derive analytical expressions for the NE takeover strategies and saddle-point values in the space of pure and mixed strategies.
   \item \textbf{Optimal linear state-feedback control policies}: For linear discrete-time dynamical system with quadratic takeover, state, and control costs, we derive an analytic state-feedback control policy coupled between the players through a scalar parameter. Compared to conventional dynamic games, we show how such a parameterization enables us to compute an analytical solution. Furthermore, we establish sufficient conditions under which the game admits a saddle-point in the space of feedback control policies that are affine in the state.
    \item \textbf{Exact takeover strategies and saddle-point value parameters for scalar system}:  
    We derive analytical state-feedback control policies of both players for a scalar linear system. In particular, we derive closed-form expressions for the NE takeover strategies and parameterized value of the game \emph{independent of the continuous state}.
    \item \textbf{Approximate takeover strategies and saddle-point value parameters for $n-$dimensional system}:  
    Using the state-feedback control policies, we derive upper and lower saddle-point value bounds for $n-$dimensional systems associated with each \texttt{FlipDyn} state.
    Using such bounds, we derive parameterized approximate NE takeover strategies and the corresponding saddle-point value. Finally, we derive conditions that characterize the difference between the approximate and true saddle-point value.
\end{enumerate}

We illustrate our results for the scalar and $n-$dimensional systems through numerical examples. 
For an $n-$dimensional system, the computational cost of the proposed method scales as $\mathcal{O}(Ln^{3})$, where $L$ denotes the finite-horizon.

This paper is organized as follows. Section~\ref{sec:Problem_Formulation} defines the general \texttt{FlipDyn} problem with arbitrary state transition dynamics and control policies under state- and control-dependent costs. 
Section~\ref{sec:FlipDyn_General_Systems} outlines a solution methodology for discrete-time dynamical systems with non-negative costs and known control policies. Section~\ref{subsec:linear_optimal_control} presents optimal linear state-feedback control policies for linear discrete-time systems with quadratic costs.
Section~\ref{subsec:FlipDyn_Con_1d} investigates takeover strategies and saddle-point value parameters for scalar systems, while Section~\ref{subsec:N_dim} extends the analysis to approximate strategies and parameters for $n-$dimensional systems. 
The paper concludes with future directions in Section~\ref{sec:Conclusion}.

\section{Problem Formulation}\label{sec:Problem_Formulation}
The common resource is described as a discrete-time dynamical system, whose state evolution is given by:
\begin{align}\label{eq:dynamics}
	x_{k+1} = F_{k}^{0}(x_k,u_k),
\end{align}
where $k$ denotes the discrete-time index, taking values from the set $\mathcal{K} := \{1,2,\dots, L\} \subset \mathbb{N}$, $x_{k} \in \mathbb{R}^{n}$ is the state of the system, $u_{k} \in \mathbb{R}^m$ is the control input of the system, and $F_{k}^{0}: \mathbb{R}^{n} \times \mathbb{R}^{m} \rightarrow \mathbb{R}^{n}$ is the state transition function.
We consider an adversary attempting to takeover the common resource. In particular, we assume the adversary to be located between the controller and actuator. 
The \texttt{FlipDyn} state, $\alpha_{k} \in \{0,1\}$ indicates whether the defender ($\alpha_k = 0$) or the adversary ($\alpha_k = 1$) has taken over the system at time $k$.
We describe a takeover action at time $k$ through $\pi^{j}_k \in \{0,1\}$, where $j = 0$ denotes the defender and $j=1$ denotes the adversary.
The binary \texttt{FlipDyn} state update based on the player's takeover action satisfies
\begin{align}\label{eq:flip_state}
	\alpha_{k+1} &= \begin{cases}
		\alpha_{k}, & \text{if } \pi^{1}_{k} = \pi^{0}_{k}, \\
		j, & \text{if } \pi^{j}_{k} = 1.
	\end{cases}
\end{align}
The \texttt{FlipDyn} state update~\eqref{eq:flip_state} indicates that if both players act to takeover the resource at the same time instant, then their actions are nullified, rendering the \texttt{FlipDyn} state to remain unchanged. 
However, if the resource is under control by one of the players, who does not exert a takeover action, while the other player attempts to takeover, then the \texttt{FlipDyn} state toggles at time $k+1$. Finally, if a player is already in control and continues the takeover while the other player remains idle, then the \texttt{FlipDyn} state remains unchanged. Thus, the \texttt{FlipDyn} dynamics is compactly described as:
\begin{align}
	\label{eq:FlipDyn_compact}
	\alpha_{k+1} &= \left(\bar{\pi}^{0}_k\bar{\pi}^{1}_k +  \pi^{0}_k\pi^{1}_k\right)\alpha_{k} + \bar{\pi}^{0}_k(\pi^{0}_k + \pi^{1}_k),
\end{align}
where a binary variable $\bar{\pi} := 1 - \pi$. Takeovers are mutually exclusive, i.e., only one player is in control of the system at any given time. The continuous state $x_{k+1}$ is a function of $\alpha_{k+1}$, modifying the state evolution~\eqref{eq:dynamics} to:
\begin{align}\label{eq:Flip_dynamics}
	x_{k+1} = (1 - \alpha_{k+1})F_{k}^{0}(x_{k},u_{k}) + \alpha_{k+1}F_{k}^{1}(x_{k},w_{k}),
\end{align}
where $F_{k}^{1}: \mathbb{R}^{n} \times \mathbb{R}^{p} \rightarrow \mathbb{R}^{n}$ is the state transition function for the adversary, and $w_k \in \mathbb{R}^{p}$ is the attack input.

In this work, we aim to design optimal control policy and  takeover strategy pairs for both player governing the described dynamical system. 
Given a non-zero initial state $x_{1}$, we pose the resource takeover and control problem as a zero-sum game goverened by the dynamics~\eqref{eq:Flip_dynamics} and~\eqref{eq:FlipDyn_compact}, over a finite-horizon $L$, where the defender aims to minimize a net cost given by:
\begin{equation}\label{eq:obj_def_alpha}
	\begin{aligned}
		& J(x_{1}, \alpha_{1}, \{\pi^1_{\mathbf{L}}\}, \{\pi^0_{\mathbf{L}}\}, u_{\mathbf{L}}^{*}, w_{\mathbf{L}}^{*}) = g_{L+1}(x_{L+1}, \alpha_{L+1}) + \\  &\sum_{t=1}^{L} g_t(x_t, \alpha_t) + \pi_{t}^{0}d_t(x_t) + \bar{\alpha}_t m_t(u_t) -\pi^{1}_{t}a_t(x_t) - \alpha_t n_t(w_t), 
	\end{aligned}
\end{equation}
where $g_t(x_t, \alpha_t): \mathbb{R}^{n} \times \{0,1\} \rightarrow \mathbb{R}$ represents the state cost with $g_{L+1}(x_{L+1}, \alpha_{L+1}): \mathbb{R}^{n} \times \{0,1\} \rightarrow \mathbb{R}$ as the terminal state cost, $d_t(x_t): \mathbb{R}^{n} \rightarrow \mathbb{R}$  and $a_t(x_t):\mathbb{R}^{n} \rightarrow \mathbb{R}$ are the instantaneous takeover costs for the defender and adversary, respectively. The terms $m_t(u_t):\mathbb{R}^{m} \rightarrow \mathbb{R}$ and $n_t(w_t): \mathbb{R}^{p} \rightarrow \mathbb{R}$ are control costs for the defender and adversary, respectively. The notations $\{\pi_{\mathbf{L}}^j\} := \{\pi_1^j, \dots, \pi_{L}^j\}, j \in \{0,1\}$, $u_{\mathbf{L}} := \{u_{1}, \dots, u_L\}$, and $w_{\mathbf{L}} := \{w_{1}, \dots, w_L\}$. The adversary aims to maximize the net cost~\eqref{eq:obj_def_alpha} leading to a zero-sum dynamic game, termed as \emph{\texttt{FlipDyn} game~\cite{FlipDyn_banik2022} with control}. 

\smallskip

We seek to find Nash Equilibria (NE) of the game~\eqref{eq:obj_def_alpha}. To guarantee the existence of a pure or mixed takeover strategy, we expand the set of player policies to behavioral strategies, i.e., probability distributions over the space of discrete actions at each time step~\cite{hespanha2017noncooperative}.
Specifically, let
\begin{equation}
	y_{k}^{\alpha_{k}} = \begin{bmatrix}
		1 - \beta_{k}^{\alpha_{k}} & \beta_{k}^{\alpha_{k}}
	\end{bmatrix}^{\text{T}} \text{ and \hfill} z_{k}^{\alpha_{k}} = \begin{bmatrix}
	1 - \gamma_{k}^{\alpha_{k}} & \gamma_{k}^{\alpha_{k}}
\end{bmatrix}^{\text{T}},
\end{equation}
be the behavioral strategies for the defender and adversary at time instant $k$ for the \texttt{FlipDyn} state $\alpha_k$, such that $\beta_{k}^{\alpha_{k}} \in [0,1]$ and $\gamma_{k}^{\alpha_{k}} \in [0,1]$, respectively. The takeover actions
\[
\pi_{k}^{0} \sim y_{k}^{\alpha_{k}}, \quad  \pi_{k}^{1} \sim z_{k}^{\alpha_{k}},  
\]
of each player at any time $k$ are sampled from the corresponding behavioral strategy. The behavioral strategies, $y_{k}^{\alpha_{k}}, z_{k}^{\alpha_{k}}  \in \Delta_{2}$, where $\Delta_{2}$ is the probability simplex in two dimensions. 
Over the finite-horizon $L$, let $y_{\mathbf{L}} := \{y_{1}^{\alpha_{1}}, y_{2}^{\alpha_{2}}, \dots, y_{L}^{\alpha_{L}}\} \in \Delta^{L}_{2}$ and $z_{\mathbf{L}} := \{z_{1}^{\alpha_{1}}, z_{2}^{\alpha_{2}}, \dots, z_{L}^{\alpha_{L}}\} \in \Delta^{L}_{2}$ be the sequence of defender and adversary behavioral strategies. Thus, the expected outcome of the zero-sum game~\eqref{eq:obj_def_alpha} is given by:
\begin{equation}
    \label{eq:opti_E_cost}
    \resizebox{0.99\linewidth}{!}{$
	J_{E}(x_1, \alpha_{1}, y_{\mathbf{L}}, z_{\mathbf{L}}, u_{\mathbf{L}}, w_{\mathbf{L}}) :=  \mathbb{E}[J( x_{1}, \alpha_{1}, \{\pi^1_{L}\}, \{\pi^0_{L}\}, u_\mathbf{L}, w_\mathbf{L})]$},
\end{equation}
where the expectation is computed with respect to the distributions $y_{\mathbf{L}}$ and $z_{\mathbf{L}}$.

\begin{definition}[Nash Equilibrium~\cite{bacsar1998dynamic}]
In a two-player zero-sum game with a payoff function $C:\Psi \times \Omega \to \mathbb{R}$, a NE is a pair of strategies \((\psi^*, \omega^*)\) for the defender and adversary, respectively, such that 
\[
C(\psi^*, \omega) \leq C(\psi^*, \omega^*) \leq C(\psi, \omega^*), \quad \forall \psi \in \Psi, \omega \in \Omega.
\]
In other words, neither player can unilaterally deviate to improve their individual payoff. \frqed
\end{definition}

In the context of the \texttt{FlipDyn-Con} framework, we seek a saddle-point solution ($y_{\mathbf{L}}^{*}, z_{\mathbf{L}}^{*}, u_{\mathbf{L}}^{*}, w_{\mathbf{L}}^{*}$) in the space of behavioral strategies and control inputs such that for any non-zero initial state $x_0 \in \mathbb{R}^{n}, \alpha_0 \in \{0,1\}$,
\begin{equation*}
    \begin{aligned}
         \underline{J}_{E} \leq J_E(x_0, \alpha_0, y_{\mathbf{L}}^{*}, z_{\mathbf{L}}^{*}, u_{\mathbf{L}}^{*}, w_{\mathbf{L}}^{*}) \leq \overline{J}_{E}, 
    \end{aligned} 
\end{equation*}
where $\underline{J}_{E}: = J_{E}(x_0, \alpha_0, y_{\mathbf{L}}^{*}, z_{\mathbf{L}},u_{\mathbf{L}}^{*}, w_{\mathbf{L}})$ and $\overline{J}_{E}:= J_{E}(x_0, \alpha_0, y_{\mathbf{L}}. z_{\mathbf{L}}^{*}, u_{\mathbf{L}}, w_{\mathbf{L}}^{*})$.
The \texttt{FlipDyn} game with control, referred to as \texttt{FlipDyn-Con}, is defined by the expected cost~\eqref{eq:opti_E_cost}, evaluated in the space of player takeover strategies and control input policies, subject to the dynamics defined in~\eqref{eq:FlipDyn_compact} and~\eqref{eq:Flip_dynamics}. 
In the next section, we will derive the takeover strategies of \texttt{FlipDyn-Con} for general systems.

\section{\texttt{FlipDyn-Con} for General Systems}\label{sec:FlipDyn_General_Systems}
We build on the FlipDyn game framework~\cite{FlipDyn_banik2022}, which models strategic mixed policy takeovers between a defender and an adversary.
In this section, we extend the \texttt{FlipDyn} model to a hybrid game-theoretic framework in which both players characterize the strategic takeovers over the space of both pure and mixed policies of a discrete-time system.

\subsection{Saddle-point value}
Given an initial \texttt{FlipDyn} state at any time instant $k \in \mathcal{K}$, the saddle-point value comprises of an instantaneous state and control cost, along with an additive cost-to-go determined by the players' takeover actions. The cost-to-go is evaluated via a cost-to-go matrix, denoted by $\Xi_{k+1}^{0} \in \mathbb{R}^{2 \times 2}$ and $\Xi_{k+1}^{1} \in \mathbb{R}^{2 \times 2}$ for the \texttt{FlipDyn} state $\alpha_{k} = 0$ and $\alpha_{k} = 1$, respectively. Let $V^{0}_k(x, u_k, \Xi_{k+1}^{0})$ and $V^{1}_k(x, w_k, \Xi_{k+1}^{1})$ denote the saddle-point values at time instant $k$, corresponding to the \texttt{FlipDyn} states $\alpha = 0$ and $\alpha = 1$, respectively, as functions of the continuous state $x$, the given control policy pair $u_k$ and $w_k$, and the associated cost-to-go matrices. The entries of the cost-to-go matrix $\Xi^{0}_{k+1}$, corresponding to each pair of takeover actions, are given by:
\begin{equation}\label{eq:Cost_to_go_al0}
    \begin{aligned}
		& \begin{matrix} & \hphantom{000} \text{Idle} & & \hphantom{v_{k+1}^0(.,.} \text{Takeover}\end{matrix} \\
		\begin{matrix} \text{Idle} \\[3pt] \text{Takeover}\end{matrix} & \underbrace{\begin{bmatrix}
			v_{k+1}^0 &  v_{k+1}^1 - a_k(x)  \\[3pt]
			v_{k+1}^0 + d_k(x) &  v_{k+1}^0 + d_k(x) - a_k(x)
		\end{bmatrix}}_{\Xi_{k+1}^{0}}
	\end{aligned},
\end{equation}
\begin{align}
    \label{eq:V_k_0} \text{where } \ & v_{k+1}^{0} := V_{k+1}^{0}\left(F_k^{0}(x,u_k),u_{k+1},\Xi_{k+2}^{0}\right), \\
    \label{eq:V_k_1} & v_{k+1}^{1} := V_{k+1}^{1}(F_k^{1}(x,w_k),w_{k+1},\Xi_{k+2}^{1}).
\end{align}
The matrix entries for $\Xi^{0}_{k+1}$ are determined using the defender and adversary control policies, and the dynamics~\eqref{eq:FlipDyn_compact} and~\eqref{eq:Flip_dynamics}. Let $X(i,j)$ corresponds to the $(i,j)$-th entry of the matrix $X$. The diagonal entries $\Xi_{k+1}^{0}(1,1)$ and $\Xi_{k+1}^{0}(2,2)$ correspond to both the defender and adversary remaining idle and taking over, respectively. The off-diagonal entries correspond to one player taking over the resource while the other remains idle. 
The cost-to-go couples the saddle-point values between the \texttt{FlipDyn} states. 
Thus, at time $k$ for a given control policy $u_k$, state $x$ and $\alpha_k=0$, the saddle-point value satisfies
\begin{equation}
    \label{eq:V_k^0_cost_to_go}
	V^{0}_k(x, u_k, \Xi_{k+1}^{0}) = g_k(x,0) + m_k(u_k) + \Val(\Xi^{0}_{k+1}), 
\end{equation}
where $\Val(X_{k+1}^{\alpha_{k}}):= \min_{y_{k}^{\alpha_{k}}} \max_{z_{k}^{\alpha_{k}}} y_{k}^{{\alpha_{k}}^{\tp}}X_{k+1}z_{k}^{\alpha_{k}}$, represents the (mixed) saddle-point value of the zero-sum matrix $X_{k+1}$ for the \texttt{FlipDyn} state $\alpha_{k}$. The defender's (row player) and adversary's (column player) action results in either an entry within $\Xi^{0}_{k+1}$ (if the matrix has a saddle point in pure strategies) or in the expected sense, resulting in a cost-to-go from state $x$ at time $k$.

Similarly, for $\alpha_k = 1$, the entries of the cost-to-go matrix $\Xi_{k+1}^{1}$ and the corresponding saddle-point value are given by:
\begin{equation}\label{eq:Cost_to_go_al1}
    \begin{aligned}
		& \begin{matrix} & \hphantom{000} \text{Idle} & & \hphantom{v_{k+1}^0} \text{Takeover}\end{matrix} \\
		\begin{matrix} \text{Idle} \\ \text{Takeover}\end{matrix} & \underbrace{\begin{bmatrix}
			v_{k+1}^{1}  &  v_{k+1}^{1} - a_k(x)  \\
			v_{k+1}^0 + d_k(x) &  v_{k+1}^{1} + d_k(x) - a_k(x)
		\end{bmatrix}}_{\Xi_{k+1}^{1}},
	\end{aligned}
\end{equation}
\begin{equation}
    \label{eq:V_k^1_cost_to_go}
	\text{with} \ V^{1}_k(x,w_k, \Xi_{k+1}^{1}) = g_k(x,1) - n_{k}(w_k) +  \Val(\Xi^{1}_{k+1}).
\end{equation}

With the saddle-point values established for each \texttt{FlipDyn} states, the following subsection characterizes the NE takeover strategies and the corresponding saddle-point values over the finite-horizon $L$.

\subsection{NE takeover strategies of the \texttt{FlipDyn} game}
To characterize the saddle-point value of the game, we impose a restriction on the cost functions, as outlined in the following mild assumption.
\begin{assumption}\label{ast:general_costs}
    [Non-negative costs] For any time instant $k \in \mathcal{K}$, the state and control-dependent costs $g_k(x,\alpha), d_k(x), a_k(x), m_k(u_{k}), n_k(w_{k}),$ for all $x \in \mathbb{R}^{n}, u_k \in \mathbb{R}^{m}, w \in \mathbb{R}^{p},$ and $\alpha \in \{0,1\}$ are non-negative $(\mathbb{R}_{\geq 0})$. 
\end{assumption}

Assumption~\ref{ast:general_costs} allows us to compare the entries of the cost-to-go matrix without altering the sign of the costs, thereby facilitating the characterization of the players' strategies (pure or mixed).
Building on this assumption, we summarize the following results, which provides a recursion of saddle-point value over the finite-horizon and the associated NE takeover strategies for both players.
To solve the \texttt{FlipDyn-Con} game, we characterize a Bellman-like dynamic programming (DP) recursion for computing the saddle-point value in the presence of adversarial takeovers.
This provides the foundation for synthesizing optimal takeover strategies.

For ease of reading, we recommend focusing first on $\alpha_{k} = 0$, where the defender is in control. 
The corresponding result for $\alpha_{k} = 1$, where the adversary controls the system, follows a similar structure and is included here for completeness.

\begin{theorem}\label{th:NE_Val_gen_FDC}
    (\underline{Case $\alpha_k = 0$}) Under Assumption~\ref{ast:general_costs}, for a fixed pair of control policies, $u_{\mathbf{L}}$ and $w_{\mathbf{L}}$, the \texttt{FlipDyn-Con} game~\eqref{eq:opti_E_cost} governed by the continuous state dynamics~\eqref{eq:Flip_dynamics} and \texttt{FlipDyn} dynamics~\eqref{eq:FlipDyn_compact}, admits a unique pair of NE takeover strategies at each time $k \in \mathcal{K}$, given by:
    \begin{align}
    \begin{split}\label{eq:def_TP_gen_FDC_al0}
            y^{0*}_{k}  = \begin{cases}
                \begin{bmatrix} \dfrac{a_k(x)}{\check{\Xi}_{k+1}} & 1 - \dfrac{a_k(x)}{\check{\Xi}_{k+1}}
            \end{bmatrix}^{\tp}, & \text{if } \ 
                \begin{matrix}
                    {\check{\Xi}_{k+1}} > d_k(x) \\
                    {\check{\Xi}_{k+1}} > a_k(x)
                \end{matrix}, 
                \\
            \begin{bmatrix} \hphantom{00} 1 & \hphantom{a_k00000} 0 \hphantom{00}
            \end{bmatrix}^{\tp}, & \text{otherwise,}
            \end{cases} 
    \end{split}\\
    \begin{split}\label{eq:adv_TP_gen_FDC_al0}
            z^{0*}_{k}  = \begin{cases}
                \begin{bmatrix} 1 - \dfrac{d_k(x)}{\check{\Xi}_{k+1}} & \dfrac{d_k(x)}{\check{\Xi}_{k+1}}
            \end{bmatrix}^{\tp}, & \text{if } \ 
                \begin{matrix}
                    {\check{\Xi}_{k+1}} > d_k(x) \\
                    {\check{\Xi}_{k+1}} > a_k(x)
                \end{matrix}, \\ 
            \begin{bmatrix} \hphantom{00} 0 & \hphantom{a_k00000} 1 \hphantom{00}
            \end{bmatrix}^{\tp},  & \text{if } \ 
                \begin{matrix}
                    {\check{\Xi}_{k+1}} \leq d_k(x) \\
                    {\check{\Xi}_{k+1}} > a_k(x)
                \end{matrix}, \\ 
            \begin{bmatrix} \hphantom{00} 1 & \hphantom{a_k00000} 0 \hphantom{00}
            \end{bmatrix}^{\tp}, & \text{otherwise,}
            \end{cases} 
    \end{split}
    \end{align}
    where ${\check{\Xi}_{k+1}} := V^{1}_{k+1}(F_k^{1}(x,w_k),w_{k+1},\Xi_{k+2}^{1}) - V^{0}_{k+1}(F_k^{0}(x,u_k),u_{k+1},\Xi_{k+2}^{0})$.
    
    The saddle-point value is given by:
    \begin{align}\label{eq:Val_gen_al0}
        v_{k}^{0} = 
        \begin{cases}
            \begin{aligned}
				& g_k(x,0) + v_{k+1}^{0} + m_k(u_k)\\[5pt] & \hphantom{g_k} + d_k(x) - \tfrac{a_k(x)d_k(x)}{\check{\Xi}_{k+1}},
            \end{aligned} &\text{if } \begin{matrix}
                    {\check{\Xi}_{k+1}} > d_k(x) \\
                    {\check{\Xi}_{k+1}} > a_k(x)
                \end{matrix}, \\
            \begin{aligned}
				& g_k(x,0) + m_k(u_k)\\[5pt] & \hphantom{g_k} + v_{k+1}^{1} - a_k(x),
            \end{aligned} &\text{if } \begin{matrix}
                    {\check{\Xi}_{k+1}} \leq d_k(x) \\
                    {\check{\Xi}_{k+1}} > a_k(x)
                \end{matrix}, \\ 
             g_k(x,0) + v_{k+1}^{0} +  m_k(u_k),
             & \text{otherwise},
		\end{cases} 
	\end{align}
    where $v_{k}^{0}:= V_{k}^{0}(x,u_{k},\Xi_{k+1}^{0})$.

    \medskip
    
    (\underline{Case $\alpha_k = 1$}) The unique NE takeover strategies are 
    \begin{align}
    \begin{split}\label{eq:def_TP_gen_FDC_al1}
            y^{1*}_{k}  = \begin{cases}
                \begin{bmatrix} 1 - \dfrac{a_k(x)}{\check{\Xi}_{k+1}} & \dfrac{a_k(x)}{\check{\Xi}_{k+1}}
            \end{bmatrix}^{\tp}, & \text{if } \ 
                \begin{matrix}
                    {\check{\Xi}_{k+1}} > d_k(x) \\
                    {\check{\Xi}_{k+1}} > a_k(x)
                \end{matrix}, \\
            \begin{bmatrix} \hphantom{00} 0 & \hphantom{a_k00000} 1 \hphantom{00}
            \end{bmatrix}^{\tp}, & \text{if } \ 
                \begin{matrix}
                    {\check{\Xi}_{k+1}} > d_k(x) \\
                    {\check{\Xi}_{k+1}} \leq a_k(x)
                \end{matrix}, \\
            \begin{bmatrix} \hphantom{00} 1 & \hphantom{a_k00000} 0 \hphantom{00}
            \end{bmatrix}^{\tp}, & \text{otherwise,}
            \end{cases} 
    \end{split}\\
    \begin{split}\label{eq:adv_TP_gen_FDC_al1}
            z^{1*}_{k}  = \begin{cases}
                \begin{bmatrix} \dfrac{d_k(x)}{\check{\Xi}_{k+1}} & 1 - \dfrac{d_k(x)}{\check{\Xi}_{k+1}}
            \end{bmatrix}^{\tp}, & \text{if } \ 
                \begin{matrix}
                    {\check{\Xi}_{k+1}} > d_k(x) \\
                    {\check{\Xi}_{k+1}} > a_k(x)
                \end{matrix}, \\
            \begin{bmatrix} \hphantom{00} 1 & \hphantom{a_k00000} 0 \hphantom{00}
            \end{bmatrix}^{\tp}, & \text{otherwise.}
            \end{cases} 
    \end{split}
    \end{align}
    The saddle-point value is given by:
    \begin{align}\label{eq:Val_gen_al1}
        v_{k}^{1} = 
        \begin{cases}
            \begin{aligned}
				& g_k(x,1) + v_{k+1}^{1} - n_k(w_k)\\[5pt] & \hphantom{g_k} - a_k(x) + \tfrac{a_k(x)d_k(x)}{\check{\Xi}_{k+1}},
            \end{aligned} &\text{if } \begin{matrix}
                    {\check{\Xi}_{k+1}} > d_k(x) \\
                    {\check{\Xi}_{k+1}} > a_k(x)
                \end{matrix}\\ 
            \begin{aligned}
				& g_k(x,1) - n_k(w_k)\\[5pt] & \hphantom{g_k} + v_{k+1}^{0}  + d_k(x),
            \end{aligned} &\text{if } \begin{matrix}
                    {\check{\Xi}_{k+1}} > d_k(x) \\
                    {\check{\Xi}_{k+1}} \leq a_k(x)
                \end{matrix}\\
             g_k(x,1) + v_{k+1}^{1} - n_k(w_k),
             & \text{otherwise},
		\end{cases} 
	\end{align}
    where $v_{k}^{1}:= V_{k}^{1}(x,w_{k},\Xi_{k+1}^{1})$. 
    The boundary condition at $k = L$ is given by:
    \begin{gather}\label{eq:b_cond_NE_Val_gen_FDC}
        u_{L+1} := \mathbf{0}_{m}, w_{L+1} := \mathbf{0}_{p}, 
        \Xi_{L+2}^{1} := \mathbf{0}_{2 \times 2}, \Xi_{L+2}^{0} := \mathbf{0}_{2 \times 2}, 
    \end{gather}
    where $\mathbf{0}_{i \times j} \in \mathbb{R}^{i \times j}$ represents a matrix of zeros.
    \frqed
\end{theorem}

The proof is provided in Appendix~\ref{app:Proof_of_Theorem_1}. 
Theorem 1 shows that the saddle-point value can be computed recursively using a \emph{one-step optimization} involving the current cost and the expected future cost-to-go. 
This mirrors standard DP but adapted to the hybrid nature of the FlipDyn-Con game.
For a finite cardinality of the state, fixed player policies $u_k$ and $w_k,  k \in \mathcal{K}$, and a finite-horizon $L$, Theorem~\ref{th:NE_Val_gen_FDC} yields an exact saddle-point value of the \texttt{FlipDyn-Con} game~\eqref{eq:opti_E_cost}. However, the computational and storage complexities scale undesirably with the cardinality of the state, especially in continuous state spaces.  To address this limitation, the next section introduces a parametric representation of the saddle-point value for linear dynamics with quadratic costs.


\section{\texttt{FlipDyn-Con} for LQ Problems}\label{sec:FlipDyn_Linear_Systems}
To address computational complexity of continuous state spaces arising in the \texttt{FlipDyn-Con} game, we restrict our attention to linear dynamical system with quadratic costs (LQ problems). Furthermore, we segment our analysis into two distinct cases: a scalar and an $n$-dimensional system. The state evolution of a linear system at any time instant $k \in \mathcal{K}$, under the defender's control satisfies:
\begin{equation}
    \label{eq:def_control_dynamics}
    \begin{aligned}
        x_{k+1} & = F_{k}^{0}(x_k, u_k) := E_kx_k + B_ku_k,
    \end{aligned}
\end{equation}
where $E_{k} \in \mathbb{R}^{n \times n}$ denotes the state transition matrix, while $B_{k} \in \mathbb{R}^{n \times m}$ represents the defender control matrix. Similarly, the dynamics of the same linear system, when controlled by the adversary satisfies:
\begin{equation}
    \label{eq:adv_control_dynamics}
    \begin{aligned}
        x_{k+1} & = F_{k}^{1}(x_k, w_k) := E_kx_k + H_kw_k,
    \end{aligned}
\end{equation}
where $H_{k} \in \mathbb{R}^{n \times p}$ denotes the adversary control matrix.
The \texttt{FlipDyn} dynamics~\eqref{eq:Flip_dynamics} then reduces to
\begin{align}\label{eq:linear_dynamics}
	x_{k+1} = E_{k}x_{k} + (1 - \alpha_{k+1})B_{k}u_{k} + {\alpha_{k+1}}H_{k}w_{k}.
\end{align}
The stage, takeover and control quadratic costs are given by:
\begin{gather}
        g_k(x,\alpha_k) := x^{\tp}G_k^{\alpha_k}x, \ d_k(x) := x^{\tp}D_kx, \ a_k(x) := x^{\tp}A_kx, \nonumber \\ \label{eq:ST_MV_cost_Q}
        m_k(u) := u^{\tp}M_ku, \quad n_k(w) := w^{\tp}N_kw,
\end{gather}
where $G_k^{\alpha_k} \in \mathbb{S}^{n \times n}_{+}, D_k \in \mathbb{S}^{n \times n}_{+}, A_k \in \mathbb{S}^{n \times n}_{+}, M_k \in \mathbb{S}^{m \times m}_{+}$ and $N_k \in \mathbb{S}^{p \times p}_{+}$ are positive definite matrices. 

\smallskip
\begin{remark}
The control policies of both players act exclusively within their respective  \texttt{FlipDyn} state. Specifically, the defender's control policy $u_k$ influences the state $x_{k+1}$ when the \texttt{FlipDyn} state is $\alpha_k = 0$, whereas the adversary's control policy $w_k$ governs $x_{k+1}$ when $\alpha_k = 1$.
\end{remark}

Given linear dynamics and quadratic costs, we will first derive the control policies for both players corresponding to the saddle-point value.

\subsection{\bf Control policy for the \texttt{FlipDyn-Con} LQ Problem}\label{subsec:linear_optimal_control}

To determine the control policies for both players, we need to solve the following problems in each of the \texttt{FlipDyn} states
\begin{equation}
    \label{eq:CProb_LQ_def}
        \min_{u_k(x)} \max_{w_k(x)} 
        \begin{cases}
            \begin{aligned}
                & v_{k+1}^{0} + u_k^{\tp}(x)M_ku_{k}(x) \\ & \hphantom{v_{k+1}^{0}} - \dfrac{x^{\tp}D_kxx^{\tp}A_kx}{\widetilde{P}_{k+1}(x)},
            \end{aligned} \ \hfill \text{if } \begin{matrix}
                \widetilde{P}_{k+1}(x) > x^{\tp}D_{k}x, \\
                \widetilde{P}_{k+1}(x) > x^{\tp}A_{k}x,
            \end{matrix} \\[20pt]
            \begin{aligned}
                & v_{k+1}^{1} + u_k^{\tp}(x)M_ku_{k}(x) \\ & \hphantom{v_{k+1}^{0}} - x_k^{\tp}A_kx ,
            \end{aligned} \ \hfill \text{if } \begin{matrix}
                \widetilde{P}_{k+1}(x) \leq x^{\tp}D_{k}x, \\
                \widetilde{P}_{k+1}(x) > x^{\tp}A_{k}x,
            \end{matrix} \\[10pt]
            v_{k+1}^{0} + u_k^{\tp}(x)M_ku_{k}(x),  \hfill \text{otherwise, and } 
        \end{cases} 
\end{equation}
\begin{equation}\label{eq:CProb_LQ_adv}
\min_{u_k(x)} \max_{w_k(x)} 
    \begin{cases}
            \begin{aligned}
                & v_{k+1}^{1} - w_k^{\tp}(x)N_kw_{k}(x) \\ & \hphantom{V_{k+1}^{0}} + \dfrac{x^{\tp}D_kxx^{\tp}A_kx}{\widetilde{P}_{k+1}(x)},
            \end{aligned} \ \hfill \text{if } \begin{matrix}
                \widetilde{P}_{k+1}(x) > x^{\tp}D_{k}x, \\
                \widetilde{P}_{k+1}(x) > x^{\tp}A_{k}x,
            \end{matrix} \\[20pt]
            \begin{aligned}
                & v_{k+1}^{0} - w_k^{\tp}(x)N_kw_{k}(x) \\ & \hphantom{V_{k+1}^{0}} + x_k^{\tp}D_kx ,
            \end{aligned} \ \hfill \text{if } \begin{matrix}
                \widetilde{P}_{k+1}(x) > x^{\tp}D_{k}x, \\
                \widetilde{P}_{k+1}(x) \leq x^{\tp}A_{k}x,
            \end{matrix} \\[10pt]
            v_{k+1}^{1} - w_k^{\tp}(x)N_kw_{k}(x),  \hfill \text{otherwise, } 
    \end{cases} 
\end{equation}
where, 
\begin{equation}\label{eq:gen_mNE_eqns}
    \begin{aligned}[b]
        & \widetilde{P}_{k+1}(x) :=  v_{k+1}^{1} - v_{k+1}^{0}.
    \end{aligned}
\end{equation}

The terms $v_{k+1}^{0}$ and $v_{k+1}^{1}$ are defined in~\eqref{eq:V_k_0} and~\eqref{eq:V_k_1}, respectively. The first condition in both~\eqref{eq:CProb_LQ_def} and~\eqref{eq:CProb_LQ_adv} pertains to NE takeover in mixed strategies by both players, while the remaining conditions correspond to playing NE takeover in pure strategies. Notably, the problems corresponding to NE takeover in mixed strategies involve the term $\widetilde{P}_{k+1}(x)$, which couples the saddle-point values between the \texttt{FlipDyn} states.
Crucially, the min-max problem corresponding to the NE takeover in pure strategies for each \texttt{FlipDyn} state depends on the solution to the NE takeover in mixed strategies $(\widetilde{P}_{k+1}(x) > x^{\tp}D_{k}x, \widetilde{P}_{k+1}(x) > x^{\tp}A_{k}x)$. Thus, we first derive the control policies for NE takeovers in mixed strategies. 
We constrain the control policies for both players to be functions of the continuous state $x$, resulting the saddle-point value for each \texttt{FlipDyn} state to depend solely on the continuous state $x$, as opposed to both the continuous state $x$ and the control input.
This restriction is formally outlined in the following assumption.

\begin{assumption}\label{ast:linear_control_space}
    We restrict the control policies to linear state-feedback functions of the continuous state $x$, defined by:
    \begin{equation}\label{eq:linear_sfdb_pol}
        u_k(x) := K_kx, \quad w_k(x) := W_kx,
    \end{equation}
    where $K_k \in \mathbb{R}^{m \times n}$ and $W_k \in \mathbb{R}^{p \times n}$ are defender and adversary control gains matrices, respectively. 
\end{assumption}

Under Assumption~\ref{ast:linear_control_space}, and based on the saddle-point values~\eqref{eq:Val_gen_al0} and~\eqref{eq:Val_gen_al1}, we propose a parametric form for the saddle-point value in each \texttt{FlipDyn} state as follows:
\begin{equation*}
    \label{eq:para_form_al_QC}
    \begin{aligned}
        & V_{k}^{0}(x, u_{k}(x),\Xi_{k+1}^{0}) \Rightarrow V_{k}^{0}(x) := x^{\tp}P^{0}_kx, \\
        & V_{k}^{1}(x, w_{k}(x), \Xi_{k+1}^{1}) \Rightarrow V_{k}^{1}(x) := x^{\tp}P^{1}_kx,
    \end{aligned}
\end{equation*}
where $P^{0}_{k} \in \mathbb{S}^{n \times n}$ and $P^{1}_{k} \in \mathbb{S}^{n \times n}$ real symmetric matrices corresponding to the \texttt{FlipDyn} states $\alpha = 0$ and $1$, respectively. We adopt Assumption~\ref{ast:linear_control_space} to factor out the state $x$ during the backward computation of the saddle-point value update. Additionally, we define a specific structure for the takeover costs, as detailed in the following assumption.
\begin{assumption}
    \label{ast:takeover_costs}
    At any time instant $k \in \mathcal{K}$, we define the defender and adversary costs as:
    \begin{equation}\label{eq:takeover_costs}
        d_k(x) := d_kx^{\tp}x, \quad a_k(x) := a_kx^{\tp}x,   
    \end{equation}
    where $d_k \in \mathbb{R}$ and $a_k \in \mathbb{R}$ are non-negative scalars. 
\end{assumption}

As shown in~\cite{FlipDyn_banik2022}, Assumption~\ref{ast:takeover_costs} plays an essential role in computing the saddle-point value for the $n$-dimensional dynamical system (Section~\ref{subsec:N_dim}). 
Next, we derive optimal state-feedback control policy pair $\{u_{k}^{*},w_{k}^{*}\}$ and establish conditions for its existence in linear systems under a mixed-strategy Nash Equilibrium (NE), expressed in a tractable closed-loop form. Here, and in the subsequent discussion, let $\mathbb{I}_{n} \in \mathbb{R}^{n \times n}$ represents the identity matrix.

\medskip

\begin{theorem}\label{th:linear_control_policy}
    Under Assumptions~\ref{ast:linear_control_space} and~\ref{ast:takeover_costs}, consider a linear dynamical system governed by~\eqref{eq:linear_dynamics}, with quadratic stage costs~\eqref{eq:ST_MV_cost_Q}, takeover costs~\eqref{eq:takeover_costs}, and \texttt{FlipDyn} dynamics~\eqref{eq:FlipDyn_compact}. 
    Then, under a mixed-strategy NE takeover for both the defender and adversary, the optimal control policy pair admits a linear state-feedback form~\eqref{eq:linear_sfdb_pol}, given by:
    \begin{equation}
        \label{eq:linear_stfb_def_mNE}
        u_{k}^{*}(x) := -\underbrace{(\hat{\eta}_{k}B_{k}^{\tp}P_{k+1}^{0}B_{k} + M_k)^{-1}(\hat{\eta}_{k}B_{k}^{\tp}P_{k+1}^{0}E_k)}_{K_k^{*}(\eta_k)}x,
    \end{equation}
        \begin{equation}
        \label{eq:linear_stfb_adv_mNE}
        w_{k}^{*}(x) := -\underbrace{(\hat{\eta}_{k}H_{k}^{\tp}P_{k+1}^{1}H_{k} - N_k)^{-1}(\hat{\eta}_{k}H_{k}^{\tp}P_{k+1}^{1}E_k)}_{W_k^{*}(\eta_k)}x,
    \end{equation}
    where $\hat{\eta}_{k}: = 1 - \eta_{k}^{2}$ and the parameter $\eta_k$ satisfies: 
    \begin{equation}\label{eq:LQ_mNE_cond1}
        \begin{aligned}
            & (E_k+H_kW_{k}^{*}(\eta_k))^{\tp}P_{k+1}^{1}(E_k+H_kW_{k}^{*}(\eta_k)) \\ &  
            - (E_k+B_kK_{k}^{*}(\eta_k))^{\tp}P_{k+1}^{0}(E_k+B_kK_{k}^{*}(\eta_k)) \succ d_k \mathbb{I}_{n},   
        \end{aligned}
    \end{equation}
    \begin{equation}\label{eq:LQ_mNE_cond2}
        \begin{aligned}
            & (E_k+H_kW_{k}^{*}(\eta_k))^{\tp}P_{k+1}^{1}(E_k+H_kW_{k}^{*}(\eta_k)) \\ &  
            - (E_k+B_kK_{k}^{*}(\eta_k))^{\tp}P_{k+1}^{0}(E_k+B_kK_{k}^{*}(\eta_k)) \succ a_k \mathbb{I}_{n},
        \end{aligned}
    \end{equation}
    \begin{equation}
        \label{eq:suff_cond_eta}
        \begin{aligned}
            & x^{\tp} \left((E_k+H_kW_{k}^{*}(\eta_k))^{\tp}P_{k+1}^{1}(E_k+H_kW_{k}^{*}(\eta_k)) - \right. \\ & \left. 
            (E_k+B_kK_{k}^{*}(\eta_k))^{\tp}P_{k+1}^{0}(E_k+B_kK_{k}^{*}(\eta_k)) \right)x = x^{\tp}x\dfrac{\sqrt{a_k d_k}}{\eta_k}.
        \end{aligned} 
    \end{equation}
    \frqed
\end{theorem}

\noindent The proof is presented in Appendix~\ref{app:Theorem_2}. Theorem~\ref{th:linear_control_policy} establishes the conditions for the existence of a linear state-feedback control policy pair.
This result shows that the optimal control policy pair can be expressed as a linear state-feedback law with a scalar gain $\eta_{k}$.
This characterization facilitates efficient computation of the saddle-point value through a backward iteration. The following result establish the bounds for the parameter $\eta_k$ associated with the mixed strategy NE takeover.

\smallskip

\begin{proposition}
    The permissible range for the parameter $\eta_{k}$, satisfying the condition in~\eqref{eq:suff_cond_eta}, is given by:
    \begin{equation}\label{eq:eta_bound}
        \begin{aligned}
            & 0 < \eta_k < \sqrt{\cfrac{\min_{\nu := \{d_{k},a_{k}\}} \nu}{\max_{\nu := \{d_{k},a_{k}\}} \nu}} < 1.
        \end{aligned}
    \end{equation} \frqed
\end{proposition}

The proof is presented in Appendix~\ref{app:Proposition_1}.
In the subsequent sections, we will illustrate how a constrained range for $\eta_k$ proves instrumental in determining a solution for both scalar and $n-$dimensional systems. 
The next result characterizes the control policy pair under mixed-strategy and pure-strategy NE takeover scenarios.

\medskip

\begin{theorem}\label{th:gen_control_policy}
    Under Assumptions~\ref{ast:linear_control_space} and~\ref{ast:takeover_costs}, consider a linear dynamical system governed by~\eqref{eq:linear_dynamics}, with quadratic costs~\eqref{eq:ST_MV_cost_Q}, takeover costs~\eqref{eq:takeover_costs}, and \texttt{FlipDyn} dynamics~\eqref{eq:FlipDyn_compact}. An optimal linear state-feedback control policy pair of the form~\eqref{eq:linear_sfdb_pol}, parameterized by a scalar $\eta_{k} \in [0,1]$ is given by:
    \begin{equation}
    \label{eq:def_control_mnpNE}
        u_k^{*}(x) =  
        \begin{cases}
            \begin{aligned}
                & K_k^{*}(\eta_k)x,
            \end{aligned} & \text{if } \begin{matrix}
                \widetilde{P}_{k+1}^{*}(x) > x^{\tp}d_k\mathbb{I}_{n}x, \\
                \widetilde{P}_{k+1}^{*}(x) > x^{\tp}a_k\mathbb{I}_{n}x,
            \end{matrix} \\[10pt]
            \begin{aligned}
                & K_k^{*}(1)x,
            \end{aligned} & \text{if } \begin{matrix}
                \widetilde{P}_{k+1}^{*}(x) \leq x^{\tp}d_k\mathbb{I}_{n}x, \\
                \widetilde{P}_{k+1}^{*}(x) > x^{\tp}a_k\mathbb{I}_{n}x,
            \end{matrix} \\[10pt]
            K_k^{*}(0)x,  & \text{otherwise, } 
        \end{cases}
    \end{equation}
    \begin{equation}
    \label{eq:adv_control_mnpNE}
        w_k^{*}(x) =  
        \begin{cases}
            \begin{aligned}
                & W_k^{*}(\eta_k)x,
            \end{aligned} & \text{if } \begin{matrix}
                \widetilde{P}_{k+1}^{*}(x) > x^{\tp}d_k\mathbb{I}_{n}x, \\
                \widetilde{P}_{k+1}^{*}(x) > x^{\tp}a_k\mathbb{I}_{n}x,
            \end{matrix} \\[10pt]
            \begin{aligned}
                & W_k^{*}(1)x,
            \end{aligned} & \text{if } \begin{matrix}
                \widetilde{P}_{k+1}^{*}(x) > x^{\tp}d_k\mathbb{I}_{n}x, \\
                \widetilde{P}_{k+1}^{*}(x) \leq x^{\tp}a_k\mathbb{I}_{n}x,
            \end{matrix} \\[10pt]
            W_k^{*}(0)x,  & \text{otherwise, } 
        \end{cases}
    \end{equation}
    where  
    \begin{equation*}
        \begin{aligned}
            &  \widetilde{P}_{k+1}^{*}(x) := x^{\tp}\left((E_{k} + H_{k}W_{k}^{*}(\eta_{k}))^{\tp}P_{k+1}^{1}(E_{k} + H_{k}W_{k}^{*}(\eta_{k})) \right.  \\ & \left. \hphantom{ \widetilde{P}_{k+1}^{*}(x)} -  
            (E_{k} + B_{k}K_{k}^{*}(\eta_{k}))^{\tp}P_{k+1}^{0}(E_{k} + B_{k}K_{k}^{*}(\eta_{k})\right)x, 
        \end{aligned}
    \end{equation*} 
such that $\eta_k, P_{k+1}^1$ and $P_{k+1}^0$ satisfy conditions~\eqref{eq:LQ_mNE_cond1}, \eqref{eq:LQ_mNE_cond2} and \eqref{eq:suff_cond_eta}.    
    \frqed
\end{theorem}

\medskip
The proof is provided in Appendix~\ref{app:Theorem_3}. 
Theorems~\ref{th:linear_control_policy} and~\ref{th:gen_control_policy} completely characterize the control policies of both players in both pure and mixed NE takeover strategies. 
This characterization enables a parameterized computation of the saddle-point value and supports the subsequent development of lower and upper bounds on the saddle-point value.
Defining the dynamics of the defender and adversary using a parameter $\zeta_{k} \in \mathbb{R}$, the continuous state evolution can be expressed as:
\begin{equation}\label{eq:linear_dynamics_LOC_DA}
    \begin{aligned}
        &x_{k+1} = \check{B}_{k}(\zeta_{k})x_{k} := (E_x + B_kK_k^{*}(\zeta_{k}))x_{k}, \\
        &x_{k+1} = \check{W}_{k}(\zeta_{k})x_{k} := (E_x + H_kW_k^{*}(\zeta_{k}))x_k.
    \end{aligned}  
\end{equation}
The parameter $\zeta_{k} = \eta_{k}$ under a mixed strategy NE takeover associated with the derived control policy pair~\eqref{eq:linear_stfb_def_mNE} and~\eqref{eq:linear_stfb_adv_mNE}. 
\textbf{Computational Costs:} The dominant cost arises from the matrix inverse operation in~\eqref{eq:linear_stfb_def_mNE} and~\eqref{eq:linear_stfb_adv_mNE}, resulting in $\mathcal{O}(m^{3})$ and $\mathcal{O}(p^{3})$. For a finite-horizon $L$, the total computation cost for determining the control policy pair is $\mathcal{O}(\text{max}(m^{3},p^{3})L)$.

Next, we outline the NE takeover strategies for both players, with the corresponding saddle-point values for each \texttt{FlipDyn} state, discrete-time linear dynamics with linear state-feedback control policies, and quadratic costs. We first analyze the scalar case, where $x$ is one-dimensional, to compute the exact saddle-point value, and then extend our analysis to approximate the saddle-point value for the $n$-dimensional case.


\subsection{\bf Scalar dynamical system}\label{subsec:FlipDyn_Con_1d}
Scalar quadratic costs any time $k \in \mathcal{K}$ associated with~\eqref{eq:ST_MV_cost_Q} are given by:
\begin{gather}
    g_k(x,\alpha_k) = G^{\alpha_k}_k x^{2}, \quad d_k(x) = d_k x^{2}, \quad a_k(x) = a_k x^{2}, \nonumber \\ \label{eq:cost_form_scalar}
    m_k(u) = M_kK_k^{2}x^{2}, \quad n_k(w) = N_kW_k^{2}x^{2},
\end{gather}
where $G^{\alpha_k}_k, d_k, a_k, M_k$ and $N_k$ are non-negative scalar parameters. For scalar  system, we use the following notation to represent the saddle-point value in each \texttt{FlipDyn} state. Let
\[V_{k}^{0}(x) := \mathbf{p}_{k}^{0}x^2, \quad V_{k}^{1}(x) := \mathbf{p}_{k}^{1}x^2,\]
where $\mathbf{p}_{k}^{\alpha} \in \mathbb{R}, \alpha \in \{0,1\}, k \in \mathcal{K}$.
Building on Theorem~\ref{th:NE_Val_gen_FDC}, we present the following result, which provides a closed-form expression of the NE takeover in both pure and mixed strategies for both players, and outlines the saddle-point value update of the parameter $\mathbf{p}_k^{\alpha}$.

\medskip

\begin{cor}\label{cor:NE_Val_1D_FDC}
    (\underline{Case $\alpha_k = 0$}) 
    The \texttt{FlipDyn-Con} game~\eqref{eq:opti_E_cost} governed by a scalar dynamical system~\eqref{eq:linear_dynamics_LOC_DA} and \texttt{FlipDyn} dynamics~\eqref{eq:FlipDyn_compact}, with quadratic costs~\eqref{eq:cost_form_scalar} and takeover costs~\eqref{eq:takeover_costs}, admits a unique pair of NE takeover strategies at each time $k \in \mathcal{K}$, given by:
    \begin{align}
    \begin{split}\label{eq:def_TP_1D_FDC_al0}
            y^{0*}_{k}  = \begin{cases}
                \begin{bmatrix} \dfrac{a_k}{\check{\mathbf{p}}_{k+1}} & 1 - \dfrac{a_k}{\check{\mathbf{p}}_{k+1}}
            \end{bmatrix}^{\tp}, & \text{if } \begin{matrix}
                {\check{\mathbf{p}}_{k+1}} > d_k, \\
                {\check{\mathbf{p}}_{k+1}} > a_k,
            \end{matrix} \\
            \begin{bmatrix} \hphantom{00} 1 & \hphantom{a_k0000} 0 \hphantom{00}
            \end{bmatrix}^{\tp}, & \text{otherwise,}
            \end{cases} 
    \end{split}\\
    \begin{split}\label{eq:adv_TP_1D_FDC_al0}
            z^{0*}_{k}  = \begin{cases}
                \begin{bmatrix} 1 - \dfrac{d_k}{\check{\mathbf{p}}_{k+1}} & \dfrac{d_k}{\check{\mathbf{p}}_{k+1}}
            \end{bmatrix}^{\tp}, & \text{if } 
                \begin{matrix}
                {\check{\mathbf{p}}_{k+1}} > d_k, \\
                {\check{\mathbf{p}}_{k+1}} > a_k,
            \end{matrix} \\
            \begin{bmatrix} \hphantom{00} 0 & \hphantom{a_k0000} 1 \hphantom{00}
            \end{bmatrix}^{\tp},  & \text{if } 
                \begin{matrix}
                {\check{\mathbf{p}}_{k+1}} \leq d_k, \\
                {\check{\mathbf{p}}_{k+1}} > a_k,
            \end{matrix} \\
            \begin{bmatrix} \hphantom{00} 1 & \hphantom{a_k0000} 0 \hphantom{00}
            \end{bmatrix}^{\tp},  & \text{otherwise,}
            \end{cases} 
    \end{split}
    \end{align}
    where
    \begin{equation*}
        \begin{aligned}
                \check{\mathbf{p}}_{k+1} := &\left(\dfrac{N_{k}^{2}\mathbf{p}_{k+1}^{1}}{(N_{k} - (1 - \eta_k^{2})H^{2}_{k}\mathbf{p}_{k+1}^{1})^{2}} - \right. \\ & \hphantom{N_k\mathbf{p}_{k+1}^{1}} \left. \dfrac{M_{k}^{2}\mathbf{p}_{k+1}^{0}}{(M_{k} + (1 - \eta_k^{2})B^{2}_{k}\mathbf{p}_{k+1}^{0})^{2}}\right)E^2_{k}.
        \end{aligned}
    \end{equation*}
    The saddle-point value parameter at time $k$ is given by:
    \begin{align}\label{eq:Val_1D_al0}
        \mathbf{p}_{k}^{0} = 
        \begin{cases}
            \begin{aligned}
				& G^{0}_k + d_k -\frac{d_k a_k}{\check{\mathbf{p}}_{k+1}} + K_{k}^{*}(\eta_{k})^{2}M_{k} \\[1pt] & + \dfrac{M_{k}^{2}\mathbf{p}_{k+1}^{0} }{(M_{k} + (1-\eta_{k}^{2})B_{k}^{2}\mathbf{p}_{k+1}^{0})^{2}}E_{k}^{2},
            \end{aligned} & \text{if } \begin{matrix}
                {\check{\mathbf{p}}_{k+1}} > d_k, \\
                {\check{\mathbf{p}}_{k+1}} > a_k,
            \end{matrix} \\[25pt]
            \begin{aligned}
				& G^{0}_k - a_k  \\[1pt] & + \dfrac{N_{k}^{2}\mathbf{p}_{k+1}^{1} }{(N_{k} - (1-\eta_{k}^{2})H_{k}^{2}\mathbf{p}_{k+1}^{1})^{2}}E_{k}^2,
            \end{aligned} & \text{if } \begin{matrix}
                {\check{\mathbf{p}}_{k+1}} \leq d_k, \\
                {\check{\mathbf{p}}_{k+1}} > a_k,
            \end{matrix} \\[25pt]
             \begin{aligned}
				& G^{0}_k + \dfrac{M_k^{2}\mathbf{p}_{k+1}^{0} }{(M_k + B^{2}_{k}\mathbf{p}_{k+1}^{0})}E_{k}^2 
            \end{aligned}
             & \text{otherwise},
		\end{cases} 
	\end{align}
    
    (\underline{Case $\alpha_k = 1$}) The unique NE takeover strategies are given by:
    \begin{align}
    \begin{split}\label{eq:def_TP_1D_FDC_al1}
            y^{1*}_{k}  = \begin{cases}
                \begin{bmatrix} 1 - \dfrac{a_k}{\check{\mathbf{p}}_{k+1}} & \dfrac{a_k}{\check{\mathbf{p}}_{k+1}}
            \end{bmatrix}^{\tp}, & \text{if } \ 
                \begin{matrix}
                {\check{\mathbf{p}}_{k+1}} > d_k, \\
                {\check{\mathbf{p}}_{k+1}} > a_k,
            \end{matrix} \\
            \begin{bmatrix} \hphantom{00} 0 & \hphantom{a_k0000} 1 \hphantom{00}
            \end{bmatrix}^{\tp}, & \text{if } \ 
                \begin{matrix}
                {\check{\mathbf{p}}_{k+1}} > d_k, \\
                {\check{\mathbf{p}}_{k+1}} \leq a_k,
            \end{matrix} \\
            \begin{bmatrix} \hphantom{00} 1 & \hphantom{a_k0000} 0 \hphantom{00}
            \end{bmatrix}^{\tp}, & \text{otherwise,}
            \end{cases} 
    \end{split}\\
    \begin{split}\label{eq:adv_TP_1D_FDC_al1}
            z^{1*}_{k}  = \begin{cases}
                \begin{bmatrix} \dfrac{d_k}{\check{\mathbf{p}}_{k+1}} & 1 - \dfrac{d_k}{\check{\mathbf{p}}_{k+1}}
            \end{bmatrix}^{\tp}, & \text{if } \ 
                \begin{matrix}
                {\check{\mathbf{p}}_{k+1}} > d_k, \\
                {\check{\mathbf{p}}_{k+1}} > a_k,
            \end{matrix} \\
            \begin{bmatrix} \hphantom{00} 1 & \hphantom{a_k0000} 0 \hphantom{00}
            \end{bmatrix}^{\tp},  & \text{otherwise,}
            \end{cases} 
    \end{split}
    \end{align}
    The saddle-point value parameter at time $k$ is given by,
    \begin{align}\label{eq:Val_1D_al1}
        \mathbf{p}_{k}^{1} = 
        \begin{cases}
            \begin{aligned}
				& G^{1}_k - a_k + \frac{d_k a_k}{\check{\mathbf{p}}_{k+1}} -W_{k}^{*}(\eta_{k})^{2}N_{k}\\[1pt] & + \dfrac{N_{k}^{2}\mathbf{p}_{k+1}^{1} }{(N_{k} - (1 - \eta_{k}^{2})H^{2}_{k}\mathbf{p}_{k+1}^{1})^{2}}E^2_{k},
            \end{aligned} & \text{if } \begin{matrix}
                {\check{\mathbf{p}}_{k+1}} > d_k, \\
                {\check{\mathbf{p}}_{k+1}} > a_k,
            \end{matrix} \\[25pt] 
            \begin{aligned}
				& G^{1}_k + d_k - \\[1pt] & + \dfrac{M_{k}^{2}\mathbf{p}_{k+1}^{0} }{(M_{k} + (1 - \eta_{k}^{2}){B}^{2}_{k}\mathbf{p}_{k+1}^{0})^{2}}E^2_{k},
            \end{aligned} & \text{if } \begin{matrix}
                {\check{\mathbf{p}}_{k+1}} > d_k, \\
                {\check{\mathbf{p}}_{k+1}} \leq a_k,
            \end{matrix} \\[25pt] 
             \begin{aligned}
				& G^{1}_k + \dfrac{N_{k}^{2}\mathbf{p}_{k+1}^{1} }{(N_{k} - H^{2}_{k}\mathbf{p}_{k+1}^{1})}E^2_{k} 
            \end{aligned}
             & \text{otherwise}.
		\end{cases} 
	\end{align}
    The recursions~\eqref{eq:Val_1D_al0} and~\eqref{eq:Val_1D_al1} hold provided,
    \begin{equation}\label{eq:suf_cond_1d}
        {(1-\eta_k^2)}\mathbf{p}_{k+1}^{0}B_{k}^{2} + M_k {>} 0, \ {(1-\eta_k^2)}\mathbf{p}_{k+1}^{1}H_{k}^{2} - N_k {<} 0.
    \end{equation}
    The terminal conditions for the recursions~\eqref{eq:Val_1D_al0} and~\eqref{eq:Val_1D_al1} are:
    \begin{equation*}
        \mathbf{p}_{L+1}^{0} := G_{L+1}^{0}, \quad \mathbf{p}_{L+1}^{1} := G_{L+1}^{1} \hfill 
    \end{equation*} \frqed
\end{cor}

The proof is presented in Appendix~\ref{app:Corollary_1}. 
Corollary~\ref{cor:NE_Val_1D_FDC} presents a closed-form solution to the \texttt{FlipDyn-Con}~\eqref{eq:opti_E_cost} game, where the NE takeover strategies are independent of continuous state. However, it is crucial to note that the saddle-point value recursion outlined in Corollary\ref{cor:NE_Val_1D_FDC} is not universally satisfied for all quadratic control costs~\eqref{eq:cost_form_scalar}. To address this, the following remark identifies the minimum adversary control cost, $N_k$, that guarantees the validity of the recursions described in~\eqref{eq:Val_1D_al0} and~\eqref{eq:Val_1D_al1}.

\smallskip

\begin{remark}\label{rem:adv_cost_p1_scalar}
For a scalar system~\eqref{eq:linear_dynamics_LOC_DA} with quadratic costs~\eqref{eq:cost_form_scalar}, the NE takeover strategies and the recursion for the saddle-point value parameter, as described in Corollary~\ref{cor:NE_Val_1D_FDC}, are guaranteed to exist if the adversary control costs $N^{*}_{k} \leq N_{k}$ satisfies
\begin{equation*}
    \begin{aligned}
        & -N^{*}_{k} + H^2_{k}\mathbf{p}_{k+1}^{1} < 0,  \forall k \in \mathcal{K}.
    \end{aligned}
\end{equation*}
\end{remark}

The parameters $N^{*}_{k}$ in Remark~\ref{rem:adv_cost_p1_scalar} can be computed using any bisection method at each time instant $k \in \mathcal{K}$.  Starting with an arbitrary adversary control cost $N_k$, the saddle-point value parameters in~\eqref{eq:Val_1D_al0} and~\eqref{eq:Val_1D_al1} are updated recursively backward in time. At each time step $k$, if the inequality $-N_{k} + H^2_{k}\mathbf{p}_{k+1}^{1} \leq 0$ is violated, the adversary cost $N_k$ is adjusted using the bisection method. This iterative process continues until the recursion converges at $k = 0$. The resulting cost $N_k^{*}$ represents the minimum adversary control cost required to maintain the validity of the saddle-point value recursion, thereby ensuring effective control of the system.

Similar to the findings in~\cite{FlipDyn_banik2022}, alongside determining the minimum adversary control costs, we can also identify a minimum adversarial state cost, $G_{k}^{1*}$, that ensures a mixed strategy NE takeover at each time step $k \in \mathcal{K}$. Such an adversarial state cost is characterized in the following remark.

\begin{remark}
    For a scalar system~\eqref{eq:linear_dynamics_LOC_DA} with quadratic costs~\eqref{eq:cost_form_scalar}, the mixed strategy NE takeover and the corresponding recursion for the saddle-point value parameter, as outlined in Corollary~\ref{cor:NE_Val_1D_FDC}, exists for an adversary state-dependent cost $G^{1*}_{k} \leq G^{1}_{k}$ provided
    \begin{equation*}
        \begin{aligned}
            & \check{\mathbf{p}}_{k+1} > d_{k}, \quad  \check{\mathbf{p}}_{k+1} > a_{k}, \quad \forall k \in \mathcal{K},
        \end{aligned}
    \end{equation*}
    with the parameters at the time $L+1$ given by:
    \begin{equation}\label{eq:final_stage_p_scalar_sys}
        \begin{aligned}
            & \mathbf{p}_{L+1}^{1} = G_{L+1}^{1*}, \quad
             \mathbf{p}_{L+1}^{0} = G_{L+1}^{0}.
        \end{aligned}
    \end{equation}
\end{remark}

\medskip

The procedure for determining the minimum state cost $G^{1*}_{k}$ is analogous to that used for $N^{*}_k$ and involves employing a bisection method. Simultaneously computing both $G^{1*}_{k}$ and $N^{*}_{k}$ requires a dual bisection approach, with an outer loop iterating for $N_{k}^{*}$ and an inner loop iterating for $G_{k}^{1*}$. This iterative process is repeated until time instant $k = 0$ is reached and convergence is achieved for both bisection methods.

\textbf{Computational Costs:} The computation of the control policy pair simplifies to $\mathcal{O}(L)$. The computation of the saddle-point value parameters of each \texttt{FlipDyn} state incurs a cost of $\mathcal{O}(1)$ per time instant.  Consequently, over a finite horizon $L$, the total computational cost is $\mathcal{O}(L)$.
Next, we illustrate the results of Corollary~\ref{cor:NE_Val_1D_FDC} through a numerical example.
\smallskip

\subsubsection*{\underline{A Numerical Example}} We evaluate the NE takeover strategies and saddle-point value parameters derived in Corollary~\ref{cor:NE_Val_1D_FDC} on a linear time-invariant (LTI) scalar system over a finite-horizon $L=20$. The quadratic costs~\eqref{eq:cost_form_scalar} are assumed to fixed $\forall k \in \mathcal{K}$, given by:
\begin{equation}\label{eq:scalar_NE_costs}
    \begin{aligned}
        G^{0}_k = G^{0} = 1, \ G^{1}_k = G^{1} & = 1, \ d_k = d = 0.45, \\
        a_k = a = 0.25, & \ M_k = M = 0.65.
    \end{aligned}
\end{equation}
The control matrices of both the players reduce to:
\begin{equation}\label{eq:scalar_NE_control_coeff}
    B_k = H_k = \Delta t, \quad  \forall k \in \mathcal{K},
\end{equation}
where $\Delta t = 0.1$. We compute the NE takeover strategies and the corresponding saddle-point value parameters for two scenarios with a fixed state transition constant $E_k = E, \forall k \in \mathcal{K}$: $E = 0.85$ and $E = 1.0$.
For $E = 0.85$, the minimal adversary control costs are:
\begin{equation}\label{eq:scalar_NE_E_l1_adv_cost}
    N^{*}_{k} = N^{*} = \begin{cases}
        0.39, & \text{if } {\check{\mathbf{p}}_{k+1}} \geq a, {\check{\mathbf{p}}_{k+1}} \geq d \\
        0.25, & \text{otherwise},
        \end{cases} 
\end{equation}
whereas for $E = 1.0$, the minimal adversary control costs are:
\begin{equation}\label{eq:scalar_NE_E_g1_adv_cost}
    N^{*}_{k} = N^{*} = \begin{cases}
        2.17, & \text{if } {\check{\mathbf{p}}_{k+1}} \geq a, {\check{\mathbf{p}}_{k+1}} \geq d, \\
        1.51, & \text{otherwise}.
    \end{cases}
\end{equation}
To obtain a mixed strategy NE takeover over the horizon $L$, we solve for adversary cost $G_k^{1*}$ for each scenario given by:
\begin{equation}\label{eq:scalar_NE_E_g1_adv_cost_G}
    G_{k}^{1*} = G^{1*} = \begin{cases}
        1.56, & \text{when } E = 0.85, \\
        1.43, & \text{when } E = 1.00.
    \end{cases}
\end{equation}

\begin{figure*}[ht]
	\begin{center}
		\subfloat[]{\includegraphics[width = 0.25\linewidth]{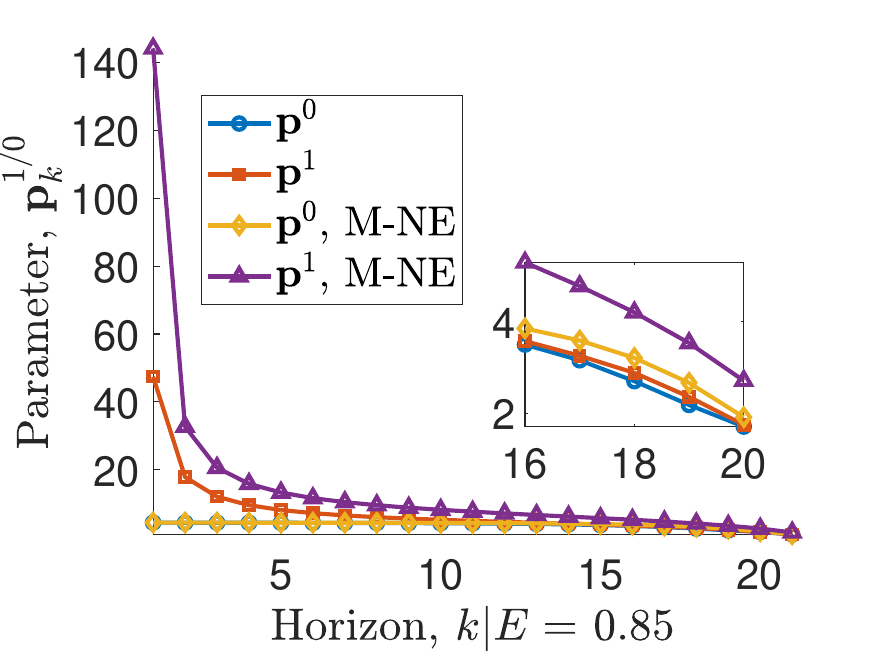}
			\label{fig:FDC_val_leq_1dim}	
		}
		\subfloat[]{\includegraphics[width = 0.25\linewidth]{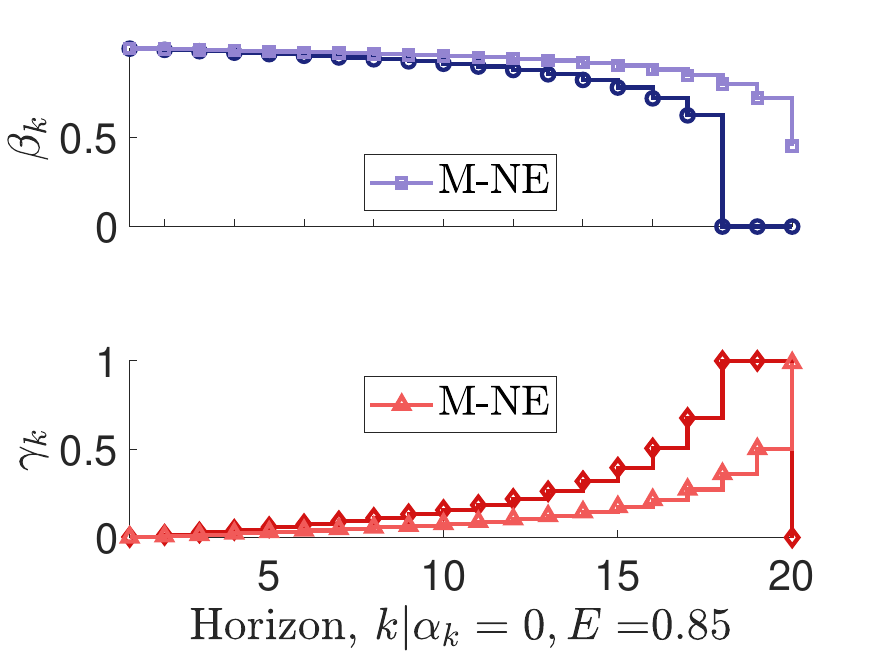}
			\label{fig:FDC_pol_leq_1dim}	
		}
		\subfloat[]{\includegraphics[width = 0.25\linewidth]{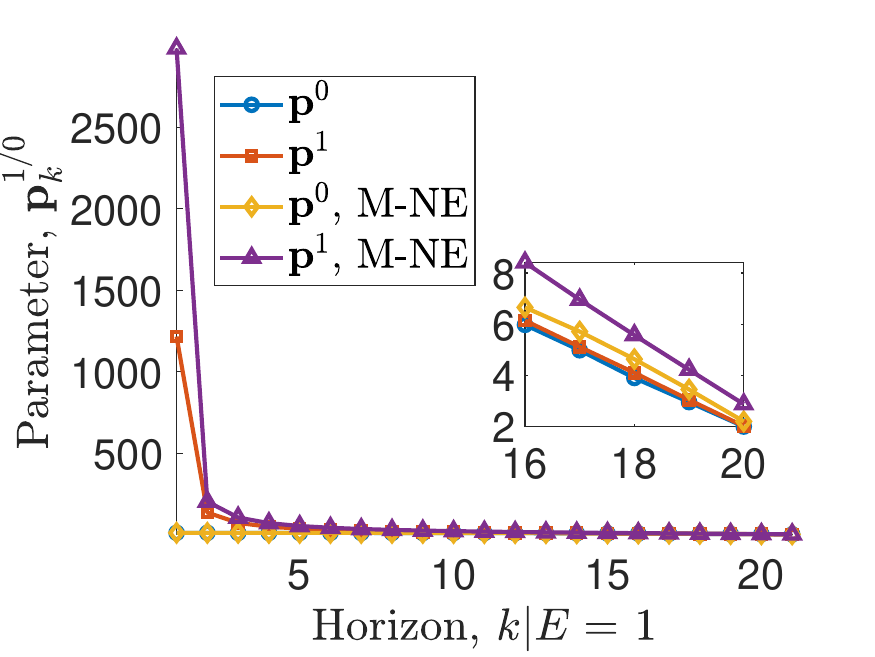}
			\label{fig:FDC_val_geq_1dim}	
		}
            \subfloat[]{\includegraphics[width = 0.25\linewidth]{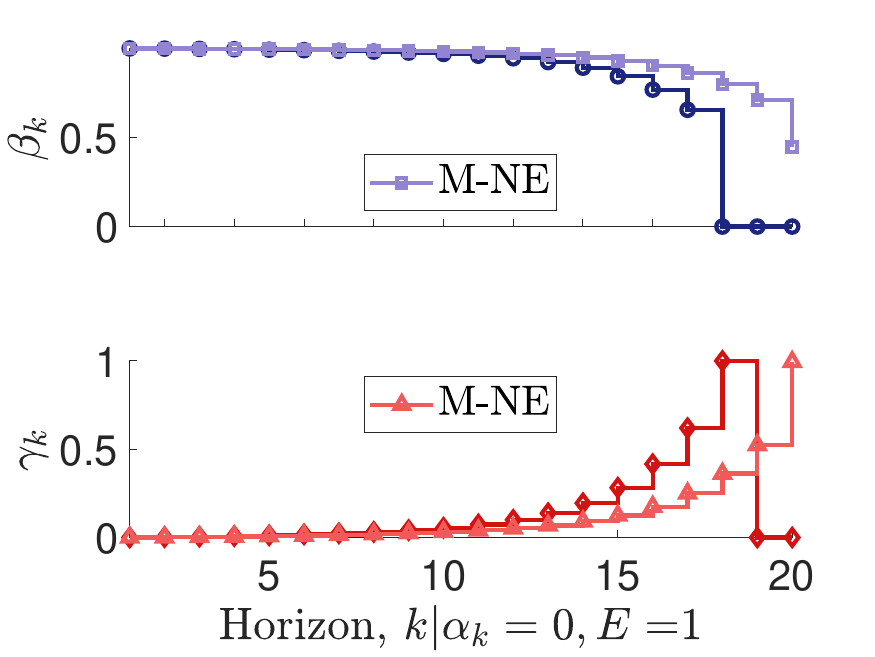}
			\label{fig:FDC_pol_geq_1dim}	
		}
    \caption{\small Saddle-point value parameters $\mathbf{p}_{k}^{i}, k \in \{1,2,\dots,L\}, i \in \{0,1\}$ for state transition constant (a) $E = 0.85$, (c) $E = 1.0$. The parameters $\mathbf{p}_{k}^{i}, $M-NE corresponds to the parameters of the saddle-point under a mixed NE takeover over the entire time horizon. Defender takeover strategies  $\beta_{k}$ and adversary takeover strategies $\gamma_{k}$ for state transition (b) $E = 0.85$ and (d) $E = 1.0$. M-NE corresponds to the mixed NE policy.}	
		\label{fig:FlipDyn_TP_CP_val}
	\end{center}
\end{figure*}

Figures~\ref{fig:FDC_val_leq_1dim} and~\ref{fig:FDC_val_geq_1dim} illustrate the saddle-point value parameters $\mathbf{p}_{k}^{0}$ and $\mathbf{p}_{k}^{1}$ for both cases: $E = 0.85$ and $1.00$. In Figure~\ref{fig:FlipDyn_TP_CP_val}, M-NE denotes a mixed strategy NE takeover spanning the entire horizon $L$, obtained using $N_{k}^{*}$ and $G_{k}^{1*}$. Notably, we observe that the saddle-point parameter value for the adversary increases with higher values of $E$, indicating that as the system transitions from open-loop stability ($E < 1$) to instability ($E \geq 1$), the adversary has a greater incentive to take control of the system. 

Figures~\ref{fig:FDC_pol_leq_1dim} and~\ref{fig:FDC_pol_leq_1dim} illustrate the takeover probabilities for the defender and adversary when \(\alpha_k = 0\). For both \(E = 0.85\) and \(E = 1.00\), the probabilities decrease (resp. increase) monotonically for the defender (resp. adversary). When the takeover strategies involve both pure and mixed strategy NE, a time instant occurs after which both players switch to pure strategies for all subsequent steps, indicating no further incentive to take over under the given costs. The difference between \(E = 0.85\) and \(E = 1.00\) highlights the rate of change in takeover strategies over time. The probability of taking over is higher for \(E = 1.00\) compared to \(E = 0.85\) but decreases sharply toward the end of the horizon.

\textbf{Comparison with LQR}: The LQR control policy~\cite{kwakernaakLinearOptimalControl1972} (Chapter 3) is a cornerstone of control theory, widely adopted for its simplicity and computational efficiency.
It arises as the extreme case of Theorem~\ref{th:gen_control_policy} when $\eta = 0$. We compare the results of Corollary~\ref{cor:NE_Val_1D_FDC} against the linear quadratic regulator (LQR) control policy, denoted as $K^{*}(0)$.
For a fair comparison, we employ the same dynamical system~\eqref{eq:scalar_NE_control_coeff} and cost structures~\eqref{eq:scalar_NE_costs},~\eqref{eq:scalar_NE_E_l1_adv_cost},~\eqref{eq:scalar_NE_E_g1_adv_cost}, and~\eqref{eq:scalar_NE_E_g1_adv_cost_G}. 
We simulate the system for 500 instances with the same initial state $x_0$ compare the resulting saddle-point value against that obtained under an LQR control policy.

\begin{figure}[ht]
\begin{center}
    \subfloat[]{\includegraphics[width = 0.5\linewidth]{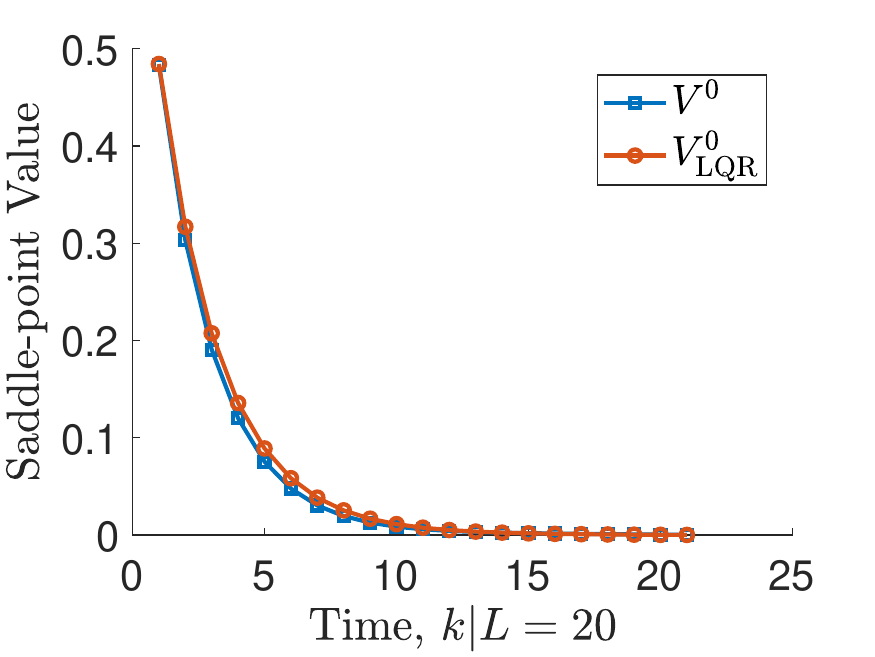}
        \label{fig:Fval_lt1_1dim}	
    }
    \subfloat[]{\includegraphics[width = 0.5\linewidth]{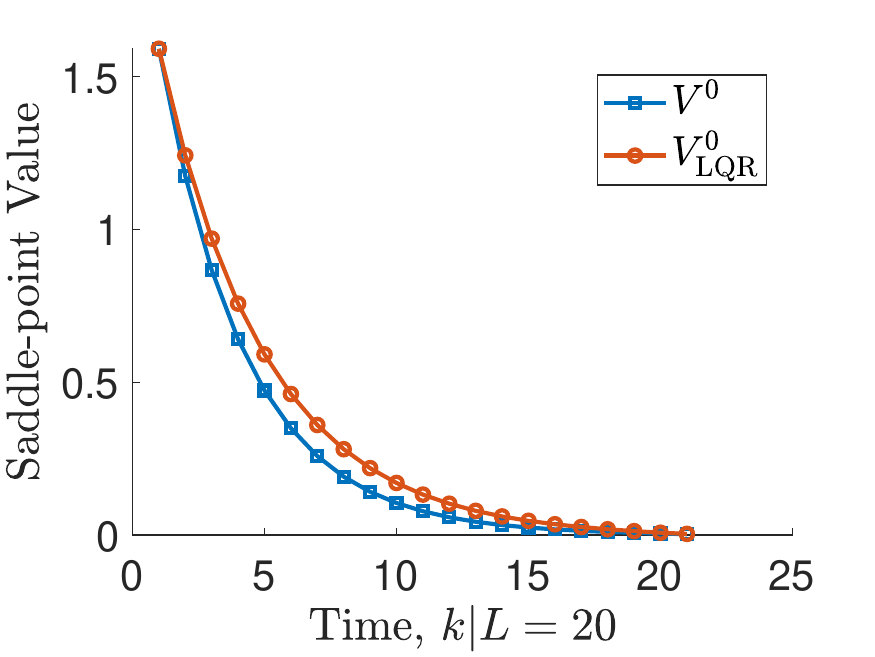}
        \label{fig:Fval_gt1_1dim}	
    }
\caption{\small Saddle-point value $V^{0}$ and $V^{0}_\text{LQR}$ (Defender LQR control law) for state transition coefficient (a) $E = 0.85$, (b) $E = 1.0$ starting with $\texttt{FlipDyn}$ state $\alpha_{0} = 0$.}	
    \label{fig:FlipDyn_SPV_compare}
\end{center}
\end{figure}
Figures~\ref{fig:Fval_lt1_1dim} and~\ref{fig:Fval_gt1_1dim}  show the saddle-point value for the initial \texttt{FlipDyn} state $\alpha_{0} = 0$, where $V^{0}$ denotes the saddle-point under the \texttt{FlipDyn-Con} game, and $V^{0}_{\text{LQR}}$ denotes the value resulting from employing an LQR defender policy. In both cases, it is clear that using the synthesized control law derived from the \texttt{FlipDyn-Con} game leads to improved performance. The results also highlight the performance loss incurred when deviating from the Nash Equilibrium strategy. 

\textbf{Threshold-based defender takeover}: We also consider a threshold-based takeover policy for the defender and compare its performance against the takeover policy derived from the \texttt{FlipDyn-Con} game.
To illustrate this, we use the same scalar system described in~\eqref{eq:scalar_NE_control_coeff}. 
We adopt the costs defined in~\eqref{eq:scalar_NE_costs} and the adversary costs specified in~\eqref{eq:scalar_NE_E_l1_adv_cost},~\eqref{eq:scalar_NE_E_g1_adv_cost}, and~\eqref{eq:scalar_NE_E_g1_adv_cost_G}. The threshold-based takeover policy is defined as follows:
\begin{align}\label{eq:def_threshold_TP}
    y^{0 : \delta}_{k}  = \begin{cases}
                    \begin{bmatrix} 0 & 1 
                    \end{bmatrix}^{\tp}, & \text{if } |x_{k}| > \delta, \\
                    \begin{bmatrix} 1 & 0 
                    \end{bmatrix}^{\tp}, & \text{otherwise}, 
                    \end{cases} 
\end{align}
where $\delta$ is heuristically set by the defender. Since the problem is formulated as a regulation task, the threshold is defined based on the absolute value of the state. We simulate the system for 500 instances with the same initial state $x_0$ and compare the resulting saddle-point value obtained under the \texttt{FlipDyn-Con} takeover policy against that achieved using the threshold-based takeover policy.
\begin{figure}[ht]
\begin{center}
    \subfloat[]{\includegraphics[width = 0.5\linewidth]{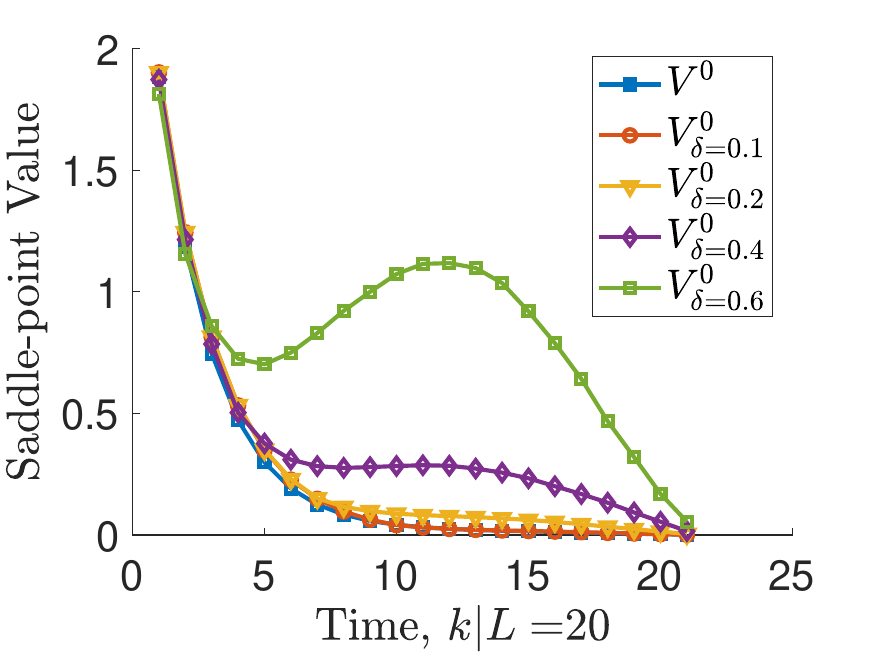}
        \label{fig:Fval_lt1_Th_1dim}	
    }
    \subfloat[]{\includegraphics[width = 0.5\linewidth]{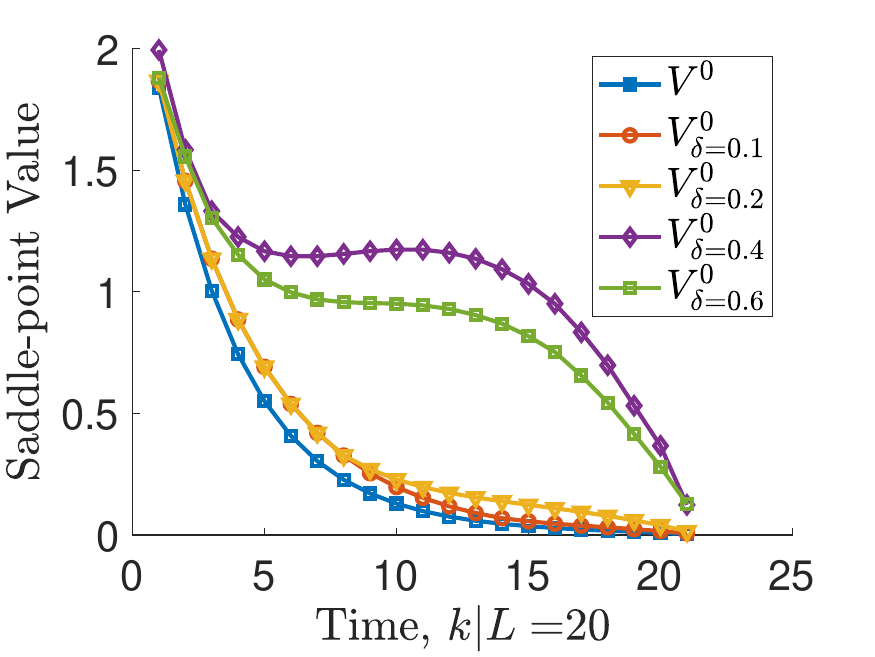}
        \label{fig:Fval_gt1_Th_1dim}	
    }
\caption{\small Saddle-point value $V^{0}$ and $V^{0}_{\delta}$ ($\delta = 0.1,0.2,0.4$ and $0.6$) for state transition coefficient (a) $E = 0.85$, (b) $E = 1.0$ starting with $\texttt{FlipDyn}$ state $\alpha_{0} = 0$.}	
    \label{fig:FlipDyn_TP_Th}
\end{center}
\end{figure}
Figures~\ref{fig:Fval_lt1_Th_1dim} and~\ref{fig:Fval_gt1_Th_1dim} show the saddle-point value for the initial state $\alpha_{0} = 0$, where $V^{0}$  denotes the saddle-point value under the \texttt{FlipDyn-Con} game and $V^{0}_{\delta}$ denotes the value obtained using the threshold-based policy with $\delta = 0.1,0.2,0.4,$ and $0.6$. 
Similar to the results with the LQR control policy, in both cases, the saddle-point value corresponding to the \texttt{FlipDyn-Con} strategy is lower than that achieved by the heuristic threshold-based takeover policy.  
Next, we extend our analysis to \(n\)-dimensional discrete-time linear dynamics with quadratic costs.

\subsection{\bf $n-$dimensional system}\label{subsec:N_dim}
Unlike the scalar case, where the state $x$ could be factored out during the computation of the mixed NE takeover strategies and saddle-point value parameters $\mathbf{p}_{k}^{0}$ and $\mathbf{p}_{k}^{1}$, such factorization does not hold for an $n-$dimensional system.  The difficulty in factoring out the state arises from the term:
\begin{equation}\label{eq:Nonlinear_Val}     
    \begin{aligned}
        \dfrac{x^{\tp}a_{k}\mathbb{I}_{n}xx^{\tp}d_{k}\mathbb{I}_{n}x}{\underbrace{x^{\tp}\left(\check{W}_{k}(\eta_{k})^{\tp}P^{1}_{k+1}\check{W}_{k}(\eta_{k}) - \check{B}_{k}(\eta_{k})^{\tp}P^{0}_{k+1}\check{B}_{k}(\eta_{k})\right)x}_{\widetilde{P}_{k+1}(x)}}.
    \end{aligned}
\end{equation}
A similar challenge was encountered in~\cite{FlipDyn_banik2022}, where the aforementioned term was approximated to factor out the state $x$ during the computation of the saddle-point value parameters. In this work, we leverage the results from Theorem~\ref{th:linear_control_policy} and propose a general approach to address such a limitation. Specifically, we utilize the parameterized control policy pair $\{u_{k}^{*}(\eta_{k}),w_{k}^{*}(\eta_{k})\}$, where the feasible parameter $\eta_k$ must satisfy the condition (\eqref{eq:suff_cond_eta}):
\begin{equation*}
    \begin{aligned}
        x^{\tp}& \left(\check{W}_{k}(\eta_{k})^{\tp}P^{1}_{k+1}\check{W}_{k}(\eta_{k}) - \right. \\ & \left. \check{B}_{k}(\eta_{k})^{\tp}P^{0}_{k+1}\check{B}_{k}(\eta_{k}) \right) x = x^{\tp}x \dfrac{\sqrt{a_{k} d_{k}}}{\eta_{k}}.
    \end{aligned}
\end{equation*}
Substituting condition~\eqref{eq:suff_cond_eta} in~\eqref{eq:Nonlinear_Val} yields:
\begin{equation}\label{eq:exact_sol_non_in_x}\dfrac{x^{\tp}a_{k}\mathbb{I}_{n}xx^{\tp}d_{k}\mathbb{I}_{n}x}{\widetilde{P}_{k+1}(x)} := \dfrac{\eta_{k} x^{\tp}a_{k}\mathbb{I}_{n}xx^{\tp}d_{k}\mathbb{I}_{n}x}{x^{\tp}x \sqrt{a_{k} d_{k}}} = \eta_{k}\sqrt{a_{k} d_{k}}x^{\tp}x.
\end{equation}
Analogous to the scalar case, we will use Theorem~\ref{th:NE_Val_gen_FDC} to present the following result, which provides a closed-form expression for the NE takeover, encompassing both both pure and mixed strategies for both players, and outlines the saddle-point value update of the parameter $P_k^{\alpha} \in \mathbb{R}^{n \times n}, \alpha \in \{0,1\}$.

\begin{cor}\label{cor:NE_Val_nD_FDC}
    (\underline{Case $\alpha_k = 0$})
    The \texttt{FlipDyn-Con} game~\eqref{eq:opti_E_cost} governed by ~\eqref{eq:linear_dynamics_LOC_DA} and \texttt{FlipDyn} dynamics~\eqref{eq:FlipDyn_compact} with quadratic costs~\eqref{eq:ST_MV_cost_Q} and takeover costs~\eqref{eq:takeover_costs}, admits a unique pair of NE takeover strategies at each time $k \in \mathcal{K}$, given by:
    \begin{align}
    \begin{split}\label{eq:def_TP_nD_FDC_al0}
            y^{0*}_{k}  = \begin{cases}
                \begin{bmatrix} \eta_{k}\sqrt{\cfrac{a_k}{d_k}} & 1 - \eta_{k}\sqrt{\cfrac{a_k}{d_k}}
            \end{bmatrix}^{\tp}, \hfill \text{if } \begin{matrix}
            \widetilde{P}_{k+1}(x) > x^{\tp}a_{k}\mathbb{I}_{n}x, \\
            \widetilde{P}_{k+1}(x) > x^{\tp}d_{k}\mathbb{I}_{n}x,
        \end{matrix}  \\
            \begin{bmatrix} \hphantom{00} 1 & \hphantom{a_k0000000} 0 \hphantom{000}
            \end{bmatrix}^{\tp}, \ \text{otherwise,}
            \end{cases} 
    \end{split}\\
    \begin{split}\label{eq:adv_TP_nD_FDC_al0}
            z^{0*}_{k}  = \begin{cases}
                \begin{bmatrix} 1 - \eta_{k}\sqrt{\cfrac{d_{k}}{a_{k}}} & \eta_{k}\sqrt{\cfrac{d_{k}}{a_{k}}}
            \end{bmatrix}^{\tp}, & \text{if } 
                \begin{matrix}
                    \widetilde{P}_{k+1}(x) > x^{\tp}a_{k}\mathbb{I}_{n}x, \\
                    \widetilde{P}_{k+1}(x) > x^{\tp}d_{k}\mathbb{I}_{n}x,
                \end{matrix}  \\
            \begin{bmatrix} \hphantom{00} 0 & \hphantom{a_k0000000} 1 \hphantom{000}
            \end{bmatrix}^{\tp},  & \text{if } 
                \begin{matrix}
                    \widetilde{P}_{k+1}(x) > x^{\tp}a_{k}\mathbb{I}_{n}x, \\
                    \widetilde{P}_{k+1}(x) \leq x^{\tp}d_{k}\mathbb{I}_{n}x,
                \end{matrix} \\
            \begin{bmatrix} \hphantom{00} 1 & \hphantom{a_k0000000} 0 \hphantom{000}
            \end{bmatrix}^{\tp},  & \text{otherwise.}
            \end{cases} 
    \end{split}
    \end{align}
    The saddle-point value parameter at time $k$ is given by:
    \begin{align}\label{eq:Val_nD_al0}
        P_{k}^{0} = 
        \begin{cases}
            \begin{aligned}
				& G^{0}_k + \check{B}_{k}(\eta_{k})^{\tp}P^{0}_{k+1}\check{B}_{k}(\eta_{k}) \\[1pt] & + K_{k}^{*}(\eta_k)^{\tp}M_kK_{k}^{*}(\eta_k) \\[1pt] & + d_k\mathbb{I}_{n} -\mathbb{I}_{n}\eta_{k}\sqrt{a_{k} d_{k}},
            \end{aligned} & \text{if } \begin{matrix}
            \widetilde{P}_{k+1}(x) > x^{\tp}a_{k}\mathbb{I}_{n}x, \\
            \widetilde{P}_{k+1}(x) > x^{\tp}d_{k}\mathbb{I}_{n}x,
        \end{matrix}  \\[15pt]
            \begin{aligned}
				& G^{0}_k + \check{W}_{k}(\eta_{k})^{\tp}{P}^{1}_{k+1}\check{W}_{k}(\eta_{k}) \\ & - a_{k}\mathbb{I}_{n},
            \end{aligned} & \text{if } \begin{matrix}
            \widetilde{P}_{k+1}(x) >  x^{\tp}a_{k}\mathbb{I}_{n}x, \\
            \widetilde{P}_{k+1}(x) \leq x^{\tp}d_{k}\mathbb{I}_{n}x,
        \end{matrix} \\[12pt]
             \begin{aligned}
				& G^{0}_k + K_{k}^{*}(0)^{\tp}M_{k}K_{k}^{*}(0) \\[1pt] & 
 + \check{B}_{k}(0)^{\tp}P^{0}_{k+1}\check{B}_{k}(0),
            \end{aligned}
             & \text{otherwise}.
		\end{cases} 
	\end{align}
    (\underline{Case $\alpha_k = 1$}) The unique NE takeover strategies are:
    \begin{align}
    \begin{split}\label{eq:def_TP_nD_FDC_al1}
            y^{1*}_{k}  = \begin{cases}
                \begin{bmatrix} 1 - \eta_{k}\sqrt{\cfrac{a_k}{d_k}} & \eta_{k}\sqrt{\cfrac{a_k}{d_k}} 
            \end{bmatrix}^{\tp}, & \text{if } \ 
                \begin{matrix}
            \widetilde{P}_{k+1}(x) > x^{\tp}a_{k}\mathbb{I}_{n}x, \\
            \widetilde{P}_{k+1}(x) > x^{\tp}d_{k}\mathbb{I}_{n}x,
        \end{matrix}  \\
            \begin{bmatrix} \hphantom{0000} 0 & \hphantom{a_k000000} 1 \hphantom{00}
            \end{bmatrix}^{\tp}, & \text{if } \ 
                \begin{matrix}
            \widetilde{P}_{k+1}(x) \leq x^{\tp}a_{k}\mathbb{I}_{n}x, \\
            \widetilde{P}_{k+1}(x) > x^{\tp}d_{k}\mathbb{I}_{n}x,
        \end{matrix} \\
            \begin{bmatrix} \hphantom{0000} 1 & \hphantom{a_k000000} 0 \hphantom{00}
            \end{bmatrix}^{\tp}, & \text{otherwise,}
            \end{cases} 
    \end{split}\\
    \begin{split}\label{eq:adv_TP_nD_FDC_al1}
            z^{1*}_{k}  = \begin{cases}
                \begin{bmatrix}  \eta_{k}\sqrt{\cfrac{d_{k}}{a_{k}}} & 1 -  \eta_{k}\sqrt{\cfrac{d_{k}}{a_{k}}}
            \end{bmatrix}^{\tp}, & \text{if } \ 
                \begin{matrix}
            \widetilde{P}_{k+1}(x) > x^{\tp}a_{k}\mathbb{I}_{n}x, \\
            \widetilde{P}_{k+1}(x) > x^{\tp}d_{k}\mathbb{I}_{n}x,
        \end{matrix}  \\
            \begin{bmatrix} \hphantom{0000} 1 & \hphantom{a_k000000} 0 \hphantom{00}
            \end{bmatrix}^{\tp},  & \text{otherwise}.
            \end{cases} 
    \end{split}
    \end{align}
    The saddle-point value parameter at time $k$ is given by,
    \begin{align}\label{eq:Val_nD_al1}
        P_{k}^{1} = 
        \begin{cases}
            \begin{aligned}
				& G^{1}_k + \check{W}_{k}(\eta_{k})^{\tp}P^{1}_{k+1}\check{W}_{k}(\eta_{k}) \\[1pt] &  - W_{k}^{*\tp}(\eta_k)N_{k}W_{k}^{*}(\eta_k) \\[1pt] &
                - a_{k}\mathbb{I}_{n} + \mathbb{I}_{n}\eta_{k}\sqrt{a_{k} d_{k}},
            \end{aligned} & \text{if } \begin{matrix}
            \widetilde{P}_{k+1}(x) > x^{\tp}a_{k}\mathbb{I}_{n}x, \\
            \widetilde{P}_{k+1}(x) > x^{\tp}d_{k}\mathbb{I}_{n}x,
        \end{matrix}  \\[15pt] 
            \begin{aligned}
				& G^{1}_k + \check{B}_{k}(\eta_{k})^{\tp}P^{0}_{k+1}\check{B}_{k}(\eta_{k}) \\ & + d_{k}\mathbb{I}_{n},
            \end{aligned} & \text{if } \begin{matrix}
            \widetilde{P}_{k+1}(x) \leq x^{\tp}a_{k}\mathbb{I}_{n}x, \\
            \widetilde{P}_{k+1}(x) > x^{\tp}d_{k}\mathbb{I}_{n}x,
        \end{matrix} \\[12pt] 
             \begin{aligned}
				& G^{1}_k - W_{k}^{*\tp}(0)N_{k}W_{k}^{*}(0) \\ & + \check{W}_{k}(0)^{\tp}P^{1}_{k+1}\check{W}_{k}(0),
            \end{aligned}
             & \text{otherwise}.
		\end{cases} 
	\end{align}
    The recursions~\eqref{eq:Val_nD_al0} and~\eqref{eq:Val_nD_al1} hold provided,
    \begin{equation}\label{eq:suf_cond_nd}
       B_{k}^{\tp}P_{k+1}^{0}B_{k} + M_k \succ 0, \ H_{k}^{\tp}P_{k+1}^{1}H_{k} - N_k \prec 0. 
    \end{equation}
    The terminal conditions for the recursions~\eqref{eq:Val_nD_al0} and~\eqref{eq:Val_nD_al1} are:
    \begin{equation*}
        P_{L+1}^{0} := G_{L+1}^{0}, \quad P_{L+1}^{1} := G_{L+1}^{1}.
    \end{equation*} \frqed
\end{cor}

The proof of Corollary~\ref{cor:NE_Val_nD_FDC} is presented in Appendix~\ref{app:Corollary_2}.
Similar to the scalar case, Corollary~\ref{cor:NE_Val_nD_FDC} provides a closed-form solution for the \texttt{FlipDyn-Con}~\eqref{eq:opti_E_cost} game with NE takeover strategies independent of state. However, such NE takeover strategies and saddle-point value parameters rely on identifying a feasible parameter $\eta_{k}, \forall k \in \mathcal{K}$, that satisfies~\eqref{eq:exact_sol_non_in_x}.
In practice, finding such a feasible $\eta_{k}$ is challenging for the linear dynamics~\eqref{eq:linear_dynamics_LOC_DA}, as the matrices $\check{B}_{k}(\zeta_{k})$ and $\check{W}_{k}(\zeta_{k})$ are generally non-diagonal. 
Therefore, there is a need to find approximate NE takeover strategies and establish bounds on the saddle-point values for general $n-$dimensional cases that may not satisfy~\eqref{eq:exact_sol_non_in_x}.
The limitation in determining a feasible $\eta_k$ is addressed by revisiting the optimal linear state-feedback control from Theorem~\ref{th:linear_control_policy}, described in the following result.

\begin{lemma}\label{lem:linear_control_policy_bounds}
    Under Assumptions~\ref{ast:linear_control_space} and~\ref{ast:takeover_costs}, consider a linear dynamical system governed by~\eqref{eq:linear_dynamics} and \texttt{FlipDyn} dynamics~\eqref{eq:FlipDyn_compact}, with quadratic costs~\eqref{eq:ST_MV_cost_Q} and takeover costs~\eqref{eq:takeover_costs}, and known saddle-point value parameters $P^{1}_{k+1}$ and $P^{0}_{k+1}$. If for every $k \in \mathcal{K}$ and $x \in \mathbb{R}^{n}$,
    \begin{equation}\label{eq:gen_SC_KW_bounds}
        B_{k}^{\tp}P_{k+1}^{0}B_{k} + M_k \succ 0, \quad H_{k}^{\tp}P_{k+1}^{1}H_{k} - N_k \prec 0,
    \end{equation}
    holds and there exist scalars $\underline{\eta}_{k} \in \mathbb{R}$ and $\overline{\eta}_{k} \in \mathbb{R}$ corresponding to an optimal linear state-feedback control pair $\{K_{k}^{*}(\underline{\eta}_{k}), W_{k}^{*}(\overline{\eta}_{k})\}$ of the form~\eqref{eq:linear_stfb_def_mNE} and~\eqref{eq:linear_stfb_adv_mNE}, such that the following conditions are satisfied: 
    \begin{equation}
        \label{eq:suff_cond_eta_bounds}
        \begin{aligned}
            x^{\tp}x\dfrac{\sqrt{a_k d_k}}{\underline{\eta}_{k}} \leq x^{\tp}\mathbf{P}_{k+1} x \leq x^{\tp}x\dfrac{\sqrt{a_k d_k}}{\overline{\eta}_{k}},
        \end{aligned} 
    \end{equation}
    \begin{equation}\label{eq:LQ_mNE_cond1_bounds}
        \begin{aligned}
            & (E_k+H_kW_{k}^{*}(\overline{\eta}_k))^{\tp}P_{k+1}^{1}(E_k+H_kW_{k}^{*}(\overline{\eta}_k)) \\ &  
            - (E_k+B_kK_{k}^{*}(\underline{\eta}_k))^{\tp}P_{k+1}^{0}(E_k+B_kK_{k}^{*}(\underline{\eta}_k)) \succ d_k \mathbb{I}_{n},   
        \end{aligned}
    \end{equation}
    \begin{equation}\label{eq:LQ_mNE_cond2_bounds}
        \begin{aligned}
            & (E_k+H_kW_{k}^{*}(\overline{\eta}_k))^{\tp}P_{k+1}^{1}(E_k+H_kW_{k}^{*}(\overline{\eta}_k)) \\ &  
            - (E_k+B_kK_{k}^{*}(\underline{\eta}_k))^{\tp}P_{k+1}^{0}(E_k+B_kK_{k}^{*}(\underline{\eta}_k)) \succ a_k \mathbb{I}_{n}.
        \end{aligned}
    \end{equation}
    where 
    \begin{equation*}
        \begin{aligned}
            \mathbf{P}_{k+1} = &(E_k+H_kW_{k}^{*}(\overline{\eta}_k))^{\tp}P_{k+1}^{1}(E_k+H_kW_{k}^{*}(\overline{\eta}_k)) 
            - \\ & (E_k+B_kK_{k}^{*}(\underline{\eta}_k))^{\tp}P_{k+1}^{0}(E_k+B_kK_{k}^{*}(\underline{\eta}_k)).
        \end{aligned}
    \end{equation*}
    Then, the saddle-point value parameters at time $k \in \mathcal{K}$, under a mixed strategy NE takeover in each \texttt{FlipDyn} state, satisfy:
    \begin{equation}
        \label{eq:P_0_ub}
         P_{k}^{0} \succeq \begin{aligned}
           & G^{0}_k + d_{k}\mathbb{I}_{n} + K_{k}^{*}(\underline{\eta}_k)^{\tp}M_kK_{k}^{*}(\underline{\eta}_k) - \mathbb{I}_{n}\underline{\eta}_{k}\sqrt{a_{k}d_{k}} \\ & + \check{B}_{k}(\underline{\eta}_{k})^{\tp}P^{0}_{k+1}\check{B}_{k}(\underline{\eta}_{k}),  
        \end{aligned}
    \end{equation}
    \begin{equation}
    \label{eq:P_1_lb}
    P_{k}^{1}  \preceq \begin{aligned}
         & G^{1}_k - a_{k}\mathbb{I}_{n} - W_{k}^{*}(\overline{\eta}_k)^{\tp}N_{k}W_{k}^{*}(\overline{\eta}_k) + \mathbb{I}_{n}\overline{\eta}_{k}\sqrt{a_{k}d_{k}} \\ &  +  \check{W}_{k}(\overline{\eta}_k)^{\tp}P^{1}_{k+1}\check{W}_{k}(\overline{\eta}_k).
        \end{aligned}
    \end{equation} 
    \frqed
\end{lemma}

The proof is derived in Appendix~\ref{app:Lemma_1}.
Lemma~\ref{lem:linear_control_policy_bounds} provides a linear state-feedback control pair that facilitates the computation of bounds on the saddle-point values independent of the state $x$, recursively backward in time. 
More importantly, condition~\eqref{eq:suff_cond_eta_bounds} serves as a relaxation for~\eqref{eq:exact_sol_non_in_x}. Such a relaxation enables us to determine an upper and lower bound in a semi-definite sense, for the saddle-point value parameters using the scalars $\overline{\eta}_{k}$ and $\underline{\eta}_{k}$.
Building on the methodology from~\cite{FlipDyn_banik2022}, we extend this approach to the $n-$dimensional case by solving for approximate NE takeover strategies and saddle-point values using the parameterization:
\begin{equation}
    \label{eq:n_dim_parameterized_spv}
    \overline{V}_{k}^{0}(x): = x^{\tp}\overline{P}_{k}^{0}x, \quad  \overline{V}_{k}^{1}(x): = x^{\tp}\overline{P}_{k}^{1}x,
\end{equation}
where $\overline{P}_{k}^{1} \in \mathbb{R}^{n \times n}$ and $\overline{P}_{k}^{0} \in \mathbb{R}^{n \times n}$.

Similar to Corollary~\ref{cor:NE_Val_nD_FDC}, we will leverage the results from Theorem~\ref{th:NE_Val_gen_FDC} to compute an approximate NE takeover pair $\{\overline{y}_{k}^{\alpha*},\overline{z}_{k}^{\alpha*}\}$, in both pure and mixed strategies of both players, and the corresponding approximate saddle-point value update of the parameter $\overline{P}_k^{\alpha} \in \mathbb{R}^{n \times n}, \alpha \in \{0,1\}$.

\begin{cor}\label{cor:Approx_NE_Val_nD_FDC}
    (\underline{Case $\alpha_k = 0$})
    
    The \texttt{FlipDyn-Con} game~\eqref{eq:opti_E_cost} governed by ~\eqref{eq:linear_dynamics_LOC_DA} and \texttt{FlipDyn} dynamics~\eqref{eq:FlipDyn_compact} with quadratic costs~\eqref{eq:ST_MV_cost_Q} and takeover costs~\eqref{eq:takeover_costs}, admits an approximate pair of NE takeover strategies at each time $k \in \mathcal{K}$, given by: 
    
    \begin{align}
    \begin{split}\label{eq:def_TP_nD_FDC_al0_bounds}
            \overline{y}^{0*}_{k}  = \begin{cases}
                \begin{bmatrix} \cfrac{a_{k}x^{\tp}x}{x^{\tp}\overline{\mathbf{P}}_{k+1}x} & 1 - \cfrac{a_{k}x^{\tp}x}{x^{\tp}\overline{\mathbf{P}}_{k+1}x}
            \end{bmatrix}^{\tp}, \ \text{if } \begin{matrix}
            \widetilde{P}_{k+1}(x) > a_{k}x^{\tp}x, \\
            \widetilde{P}_{k+1}(x) > d_{k}x^{\tp}x,
        \end{matrix}  \\
            \begin{bmatrix} \hphantom{00} 1 & \hphantom{a_k0000000000} 0 \hphantom{00...}
            \end{bmatrix}^{\tp}, \ \text{otherwise,}
            \end{cases} 
    \end{split}\\
    \begin{split}\label{eq:adv_TP_nD_FDC_al0_bounds}
            \overline{z}^{0*}_{k}  = \begin{cases}
                \begin{bmatrix} 1 - \cfrac{d_{k}x^{\tp}x}{x^{\tp}\overline{\mathbf{P}}_{k+1}x} & \cfrac{d_{k}x^{\tp}x}{x^{\tp}\overline{\mathbf{P}}_{k+1}x}
            \end{bmatrix}^{\tp}, \ \text{if } 
                \begin{matrix}
                    \widetilde{P}_{k+1}(x) > a_{k}x^{\tp}x, \\
                    \widetilde{P}_{k+1}(x) > d_{k}x^{\tp}x,
                \end{matrix}  \\
            \begin{bmatrix} \hphantom{0000} 0 & \hphantom{a_k00000000} 1 \hphantom{00...}
            \end{bmatrix}^{\tp}, \ \text{if } 
                \begin{matrix}
                    \widetilde{P}_{k+1}(x) > a_{k}x^{\tp}x, \\
                    \widetilde{P}_{k+1}(x) \leq d_{k}x^{\tp}x,
                \end{matrix} \\
            \begin{bmatrix} \hphantom{0000} 1 & \hphantom{a_k00000000} 0 \hphantom{00...}
            \end{bmatrix}^{\tp},  \ \text{otherwise,}
            \end{cases} 
    \end{split}
    \end{align}
    where 
    \[
    \overline{\mathbf{P}}_{k+1} := \check{W}_{k}(\overline{\eta}_k)^{\tp}\overline{P}^{1}_{k+1}\check{W}_{k}(\overline{\eta}_k) - \check{B}_{k}(\underline{\eta}_k)^{\tp}\overline{P}^{0}_{k+1}\check{B}_{k}(\underline{\eta}_k),
    \]
    and $\widetilde{P}_{k+1}(x)  := x^{\tp}\overline{\mathbf{P}}_{k+1}x$.
    
    The approximate saddle-point value parameter at time $k$ is given by:
    \begin{align}\label{eq:Val_nD_al0_bounds}
        \overline{P}_{k}^{0} = 
        \begin{cases}
            \begin{aligned}
				& G^{0}_k + \check{B}_{k}^{\tp}(\underline{\eta}_{k})\overline{P}^{0}_{k+1}\check{B}_{k}(\underline{\eta}_{k}) \\[1pt] & + K_{k}^{*\tp}(\underline{\eta}_{k})M_kK_{k}^{*}(\underline{\eta}_{k}) \\[1pt] & + d_k\mathbb{I}_{n} -\mathbb{I}_{n}\underline{\eta}_{k}\sqrt{a_{k} d_{k}},
            \end{aligned} \ \text{if } \begin{matrix}
            \widetilde{P}_{k+1}(x) > a_{k}x^{\tp}x, \\
            \widetilde{P}_{k+1}(x) > d_{k}x^{\tp}x,
        \end{matrix}  \\[15pt]
            \begin{aligned}
				& G^{0}_k + \check{W}_{k}^{\tp}(\overline{\eta}_k)\overline{P}^{1}_{k+1}\check{W}_{k}(\overline{\eta}_k) \\ & - a_{k}\mathbb{I}_{n},
            \end{aligned} \ \text{if } \begin{matrix}
            \widetilde{P}_{k+1}(x) >  a_{k}x^{\tp}x, \\
            \widetilde{P}_{k+1}(x) \leq d_{k}x^{\tp}x,
        \end{matrix} \\[15pt]
             \begin{aligned}
				& G^{0}_k + K_{k}^{*\tp}(0)M_{k}K_{k}^{*}(0) \\[1pt] & 
 + \check{B}_{k}^{\tp}(0)\overline{P}^{0}_{k+1}\check{B}_{k}(0),
            \end{aligned}
             \ \qquad \text{otherwise}.
		\end{cases} 
	\end{align}
    (\underline{Case $\alpha_k = 1$}) The approximate NE takeover strategies are given by:
    \begin{align}
    \begin{split}\label{eq:def_TP_nD_FDC_al1_bounds}
            \overline{y}^{1*}_{k}  = \begin{cases}
                \begin{bmatrix} 1 -  \cfrac{a_{k}x^{\tp}x}{x^{\tp}\overline{\mathbf{P}}_{k+1}x} &  \cfrac{a_{k}x^{\tp}x}{x^{\tp}\overline{\mathbf{P}}_{k+1}x}
            \end{bmatrix}^{\tp},  \text{if } \ 
                \begin{matrix}
            \widetilde{P}_{k+1}(x) > a_{k}x^{\tp}x, \\
            \widetilde{P}_{k+1}(x) > d_{k}x^{\tp}x,
        \end{matrix}  \\
            \begin{bmatrix} \hphantom{000000} 0 & \hphantom{a_k0000000} 1 \hphantom{00.}
            \end{bmatrix}^{\tp}, \text{if } \ 
                \begin{matrix}
            \widetilde{P}_{k+1}(x) \leq a_{k}x^{\tp}x, \\
            \widetilde{P}_{k+1}(x) > d_{k}x^{\tp}x,
        \end{matrix} \\
            \begin{bmatrix} \hphantom{000000} 1 & \hphantom{a_k0000000} 0 \hphantom{00.}
            \end{bmatrix}^{\tp}, \ \text{otherwise,}
            \end{cases} 
    \end{split}
    \end{align}
    \begin{align}
    \begin{split}\label{eq:adv_TP_nD_FDC_al1_bounds}
            \overline{z}^{1*}_{k}  = \begin{cases}
                \begin{bmatrix}   \dfrac{d_{k}x^{\tp}x}{x^{\tp}\overline{\mathbf{P}}_{k+1}x} & 1 -  \dfrac{d_{k}x^{\tp}x}{x^{\tp}\overline{\mathbf{P}}_{k+1}x}
            \end{bmatrix}^{\tp}, \text{if } \ 
                \begin{matrix}
            \widetilde{P}_{k+1}(x) > a_{k}x^{\tp}x, \\
            \widetilde{P}_{k+1}(x) > d_{k}x^{\tp}x,
        \end{matrix}  \\
            \begin{bmatrix} \hphantom{0000} 1 & \hphantom{a_k000000000} 0 \hphantom{00..}
            \end{bmatrix}^{\tp},  \ \text{otherwise.}
            \end{cases} 
    \end{split}
    \end{align}
    The approximate saddle-point value parameter at time $k$ is given by,
    \begin{align}\label{eq:Val_nD_al1_bounds}
        \overline{P}_{k}^{1} = 
        \begin{cases}
            \begin{aligned}
				& G^{1}_k + \check{W}_{k}^{\tp}(\overline{\eta}_k)\overline{P}^{1}_{k+1}\check{W}_{k}(\overline{\eta}_k) \\[1pt] &  - W_{k}^{*\tp}(\overline{\eta}_k)N_{k}W_{k}^{*}(\overline{\eta}_k) \\[1pt] &
                - a_{k}\mathbb{I}_{n} + \mathbb{I}_{n}\overline{\eta}_k\sqrt{a_{k} d_{k}},
            \end{aligned} \ \text{if } \begin{matrix}
            \widetilde{P}_{k+1}(x) > a_{k}x^{\tp}x, \\
            \widetilde{P}_{k+1}(x) > d_{k}x^{\tp}x,
        \end{matrix}  \\[10pt] 
            \begin{aligned}
				& G^{1}_k + \check{B}_{k}^{\tp}(\underline{\eta}_k)\overline{P}^{0}_{k+1}\check{B}_{k}(\underline{\eta}_k) \\ & + d_{k}\mathbb{I}_{n},
            \end{aligned} \ \text{if } \begin{matrix}
            \widetilde{P}_{k+1}(x) \leq a_{k}x^{\tp}x, \\
            \widetilde{P}_{k+1}(x) > d_{k}x^{\tp}x,
        \end{matrix} \\[10pt] 
             \begin{aligned}
				& G^{1}_k - W_{k}^{*\tp}(0)N_{k}W_{k}^{*}(0) \\ & + \check{W}_{k}^{\tp}(0)\overline{P}^{1}_{k+1}\check{W}_{k}(0),
            \end{aligned}
             \ \text{otherwise}.
		\end{cases} 
	\end{align}
    The recursions~\eqref{eq:Val_nD_al0_bounds} and~\eqref{eq:Val_nD_al1_bounds} hold provided,
    \begin{equation}\label{eq:suf_cond_nd_bounds}
        B_{k}^{\tp}\overline{P}_{k+1}^{0}B_{k} + M_k \succ 0, \ H_{k}^{\tp}\overline{P}_{k+1}^{1}H_{k} - N_k \prec 0.
    \end{equation}
    The terminal conditions for the recursions~\eqref{eq:Val_nD_al0_bounds} and~\eqref{eq:Val_nD_al1_bounds} are:
    \begin{equation*}
        \overline{P}_{L+1}^{0} := G_{L+1}^{0}, \quad \overline{P}_{L+1}^{1} := G_{L+1}^{1}.
    \end{equation*} \frqed
\end{cor}

Recursions~\eqref{eq:Val_nD_al0_bounds} and~\eqref{eq:Val_nD_al1_bounds} provides an approximate saddle-point parameter update. Analogous to the parameter $\eta_{k}$ range established in Lemma~\ref{lem:linear_control_policy_bounds}, the parameters $\overline{\eta}_{k}$ and $\underline{\eta}_{k}$ for a mixed strategy NE takeover can be bounded using condition~\eqref{eq:suff_cond_eta}, as detailed in the following remark.

\begin{remark}\label{rem:eta_multiple_bounds}
    The permissible range for the parameters $\overline{\eta}_{k}$ and $\underline{\eta}_{k}$ satisfying  condition~\eqref{eq:suff_cond_eta_bounds} corresponding to a mixed strategy NE is given by:
    \begin{equation}\label{eq:eta_multiple_bound}
        \begin{aligned}
            & 0 < \overline{\eta}_{k} \leq \eta_k \leq \underline{\eta}_{k} < \sqrt{\dfrac{\min_{\nu := \{d_{k},a_{k}\}} \nu}{\max_{\nu := \{d_{k},a_{k}\}} \nu}} < 1.
        \end{aligned}
    \end{equation} 
\end{remark}
\medskip
Remark~\ref{rem:eta_multiple_bounds} directly follows from Lemma~\ref{lem:linear_control_policy_bounds}.
As with the scalar case, not all control costs~\eqref{eq:ST_MV_cost_Q} satisfy the approximate  saddle-point recursion. The following remark identifies the minimum adversarial control cost required to satisfy the recursions~\eqref{eq:Val_nD_al0_bounds} and~\eqref{eq:Val_nD_al1_bounds}.

\begin{remark}\label{rem:adv_cost_approx_P}
For an $n-$dimensional system~\eqref{eq:linear_dynamics_LOC_DA} with quadratic costs~\eqref{eq:cost_form_scalar}, the NE takeover strategies and the saddle-point value parameter recursion, as outlined in Corollary~\ref{cor:Approx_NE_Val_nD_FDC}, exist for an adversary control costs $N^{*}_{k} {\prec} N_{k}$ provided the following condition holds: 
\begin{equation*}
    \begin{aligned}
        & -N^{*}_{k} + H^{\tp}_{k}\overline{P}_{k+1}^{1}H_{k} {\prec} \, 0,  \quad \forall k \in \mathcal{K}.
    \end{aligned}
\end{equation*}
\end{remark}
\medskip
Analogous to the scalar case, the parameter $N^{*}_{k}$ can be found using a bisection method $\forall k \in \mathcal{K}$. An initial candidate value can be set to $N_{L} := \nu_{\mathbb{R}_{> 0}}\mathbb{I}_{p}$, such that $\nu\mathbb{I}_{p} \succ H^{\tp}_{L}\overline{P}_{L+1}^{1}H_{L}$. Similarly, a minimum adversarial state cost $G_{k}^{1*}$ can be computed to ensure a mixed strategy NE takeover at every time step $k \in \mathcal{K}$ for the $n-$dimensional system.  The following remark summarizes such an adversarial cost.

\begin{remark}\label{rem:adv_cost_G_approx_P}
For an $n-$dimensional system~\eqref{eq:linear_dynamics_LOC_DA} with quadratic costs~\eqref{eq:cost_form_scalar}, the NE takeover strategies and the saddle-point value parameter recursion, as outlined in Corollary~\ref{cor:Approx_NE_Val_nD_FDC}, exist for an adversary state-dependent cost $G^{1*}_{k} \preceq G^{1}_{k}$ provided the following condition holds: 
    \begin{equation*}
        \begin{aligned}
            & {\overline{\mathbf{P}}}_{k+1} \succ d_{k}\mathbb{I}_{n}, \quad  {\overline{\mathbf{P}}}_{k+1} \succ a_{k}\mathbb{I}_{n}, \quad \forall k \in \mathcal{K},
        \end{aligned}
    \end{equation*}
    with the saddle-point value parameters at time $L+1$ given by:
    \begin{equation}\label{eq:final_stage_p_N_sys}
        \begin{aligned}
            & \overline{P}_{L+1}^{1} := G_{L+1}^{1*}, \
             \overline{P}_{L+1}^{0} := G_{L+1}^{0}.
        \end{aligned}
    \end{equation}
\end{remark}
\medskip
As in the scalar case, the parameter $G^{1*}_{k}$ can be determined using a bisection method. Furthermore, both $G^{1*}_{k}$ and $N^{*}_k$ can be simultaneously computed using a double bisection method.

\begin{cor}\label{cor:Approx_error}
    The error between the true and approximate saddle-point value is zero under the condition:
    
    \begin{equation}\label{eq:approx_error_cond}
        \overline{\eta}_{k} = \underline{\eta}_{k} = \gamma_{k}^{\alpha_{k}}\sqrt{\frac{a_{k}}{d_{k}}} = \left(1 - \beta_{k}^{\alpha_{k}}\right)\sqrt{\dfrac{d_{k}}{a_{k}}}.
    \end{equation}
    
    The error between the bounds is given by:
    \[
    \left(\underline{\eta}_{k} - \overline{\eta}_{k}\right)\sqrt{a_{k}d_{k}}x^{\tp}x.
    \]
\end{cor}

\smallskip

We omit the proof for Corollary~\ref{cor:Approx_error}, as~\eqref{eq:approx_error_cond} is derived by taking the difference between $P_{k}^{\alpha_{k}}$ and $\overline{P}_{k}^{\alpha_{k}}$. However, since \(\gamma_{k}^{\alpha_{k}}\) depends on \(\overline{\eta}_{k}\) and \(\underline{\eta}_{k}\), finding a feasible solution to satisfy such a condition is not always practical.
The condition~\eqref{eq:approx_error_cond} represents an equilibrium where the transitions (due to takeovers) are weighed with their respective costs, resulting in no discrepancy between approximate and true value functions.

\textbf{Computational Costs:} The control policy pair requires  $\mathcal{O}(\text{max}(m^{3},p^{3})L)$ operations. The computation of the saddle-point parameters incurs a cost of $\mathcal{O}(n^3)$ per time instant. Consequently, over a finite horizon $L$, the total computational cost amounts to $\mathcal{O}(n^3)$ + $\mathcal{O}(\text{max}(m^{3},p^{3})L)$.


\begin{figure*}[ht]
	\begin{center}
		\subfloat[]{\includegraphics[width = 0.25\linewidth]{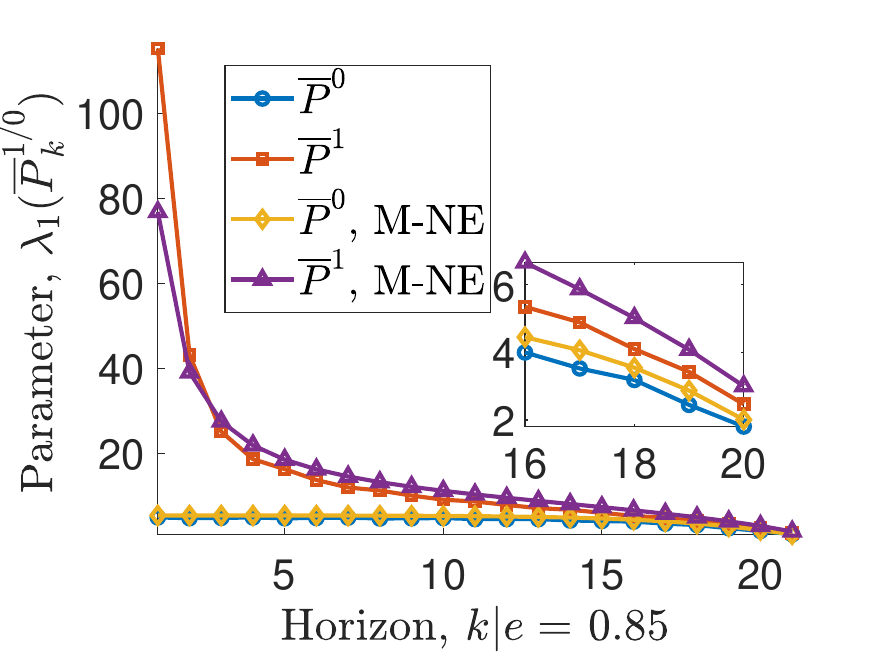}
			\label{fig:FDC_val_leq_ndim}	
		}
		\subfloat[]{\includegraphics[width = 0.25\linewidth]{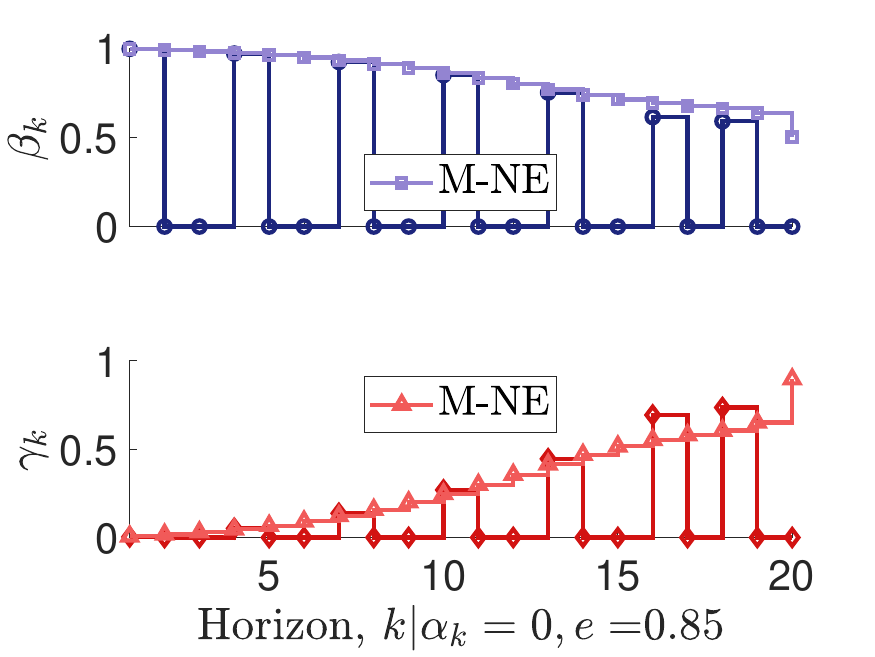}
			\label{fig:FDC_pol_leq_ndim}	
		}
		\subfloat[]{\includegraphics[width = 0.25\linewidth]{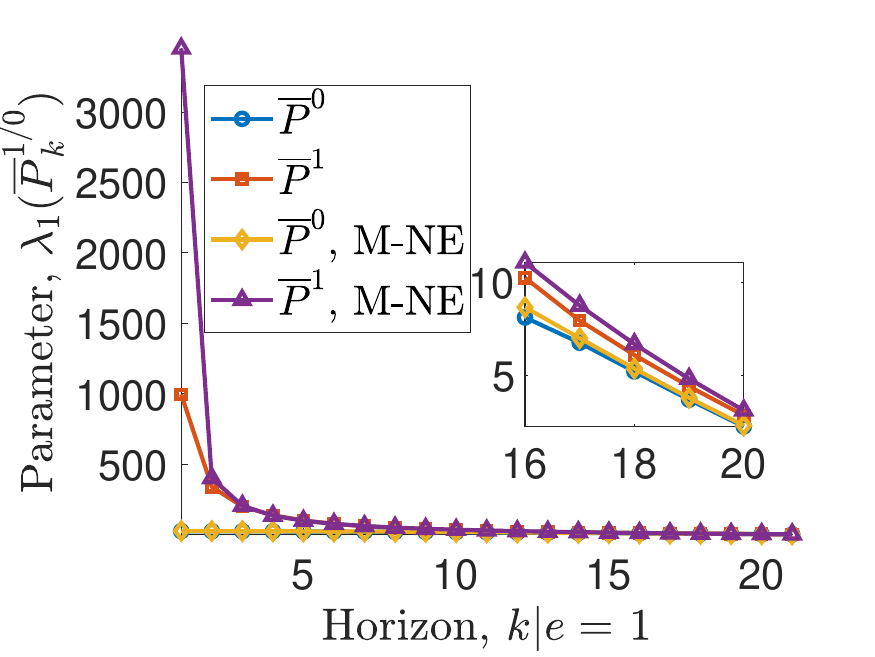}
			\label{fig:FDC_val_geq_ndim}	
		}
            \subfloat[]{\includegraphics[width = 0.25\linewidth]{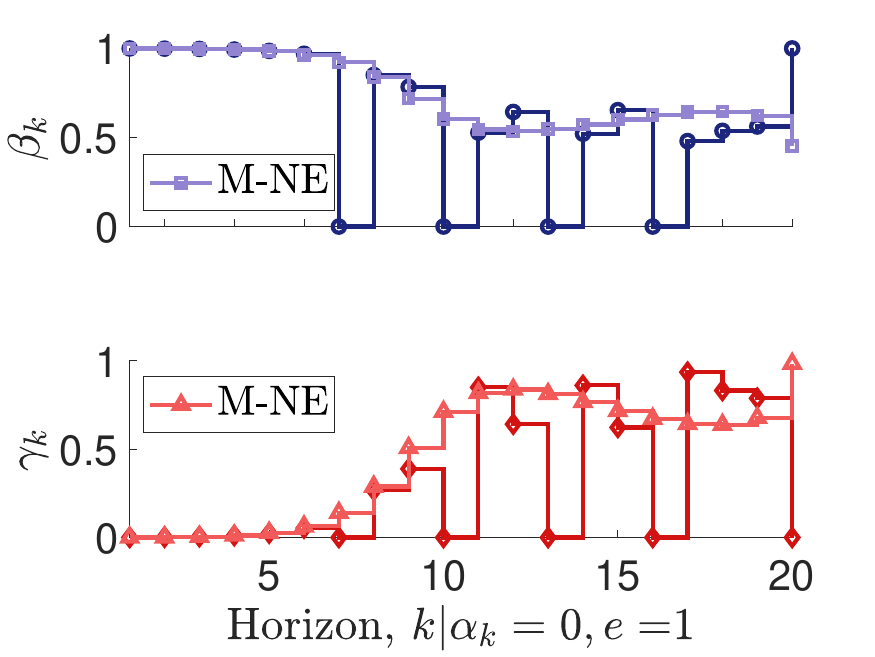}
			\label{fig:FDC_pol_geq_ndim}	
		}
    \caption{\small Maximum eigenvalues ($\lambda_{1}(\overline{P}_{k}^{\alpha})$) of saddle-point value parameters $\overline{P}_{k}^{\alpha}, k \in \{0,1,\dots,L+1\}, \alpha \in \{0,1\}$ for state transition constant (a) $e = 0.85$, (c) $e = 1.0$. The parameters $P_{k}^{i}, $M-NE corresponds to saddle-point value parameter recursion under a mixed NE takeover over the entire time horizon. Defender takeover strategy $\beta_{k}$ and adversary takeover strategy $\gamma_{k}$ for state transition (b) $e = 0.85$ and (d) $e = 1.0$. M-NE corresponds to the mixed NE policy.}	
		\label{fig:FlipDyn_TP_CP_val_ndim}
	\end{center}
\end{figure*}

\subsubsection*{\underline{A Numerical Example}} We now evaluate the results of  Corollary~\ref{cor:Approx_NE_Val_nD_FDC}, on a discrete-time two-dimensional linear time-invariant system (LTI) for a horizon length of $L=20$. The quadratic costs~\eqref{eq:ST_MV_cost_Q} are assumed to be fixed $\forall k \in \mathcal{K}$, and are given by:
\begin{equation*}
    \begin{aligned}
        G^{0}_k = G^{0} = \mathbb{I}_{n}, \ G^{1}_k = G^{1} = & 1.35\mathbb{I}_{n},  \ D_k = D = 0.45\mathbb{I}_{n}, \\
        A_k = A = 0.25\mathbb{I}_{n}, & \ M_k = M = 0.65.
    \end{aligned}
\end{equation*}
The system transition matrix $E_k = E$ and control matrices for the defender and adversary are given by:
\[
E_k = E = \begin{bmatrix}
            e & \Delta t \\ 0 & e 
            \end{bmatrix}, 
B_k = H_k = \begin{bmatrix}
            \Delta t \\ 0
            \end{bmatrix}, \quad  \forall k \in \mathcal{K},
\]
where $\Delta t = 0.1$. Similar to the scalar case, we solve for the approximate NE takeover strategies and saddle-point value parameters for two scenarios with a fixed state transition constant $e_{k} = e, \forall k \in \mathcal{K}$: $e = 0.85$ and $1.0$. Since the saddle-point value parameters for $n$-dimensions are symmetric positive definite matrices, we plot the maximum eigenvalues of the matrices $\overline{P}_{k}^{1}, \overline{P}_{k}^{0}$ in Figure~\ref{fig:FDC_val_leq_ndim} and~\ref{fig:FDC_val_geq_ndim}, respectively. In these figures, M-NE represents a mixed strategy NE takeover spanning the entire horizon $L$, obtained using $N^{*}$ and $G^{1*}$. For the case of $e = 0.85$, the costs $N_{k}^{*}, \forall k \in \mathcal{K}$, are given by:
\begin{equation*}
    N_{k}^{*} = N^{*} = \begin{cases}
        0.42, & \text{if } \begin{matrix}
            \overline{P}_{k+1}(x) \geq a_{k}x^{\tp}x, \\ \overline{P}_{k+1}(x) \geq d_{k}x^{\tp}x,
        \end{matrix} \\
        0.45, & \text{otherwise },
        \end{cases} 
\end{equation*}
and for the case of $e = 1.0$:
\begin{equation*}
    N_{k}^{*} = N^{*} = \begin{cases}
        3.73, & \text{if }\begin{matrix}
            \overline{P}_{k+1}(x) \geq a_{k}x^{\tp}x, \\ \overline{P}_{k+1}(x) \geq d_{k}x^{\tp}x,
        \end{matrix} \\
        3.40, & \text{otherwise. }
        \end{cases} 
\end{equation*}
Similarly, the minimum adversarial state cost $G_k^{1*}$ for each case of $e$, which corresponds to a mixed strategy NE takeover spanning the entire time horizon $L$, is given by:
\begin{equation*}
    G^{1*}_{k} = G^{1*} = \begin{cases}
        1.67\mathbb{I}_{n}, & \text{when } e = 0.85, \\
        1.48\mathbb{I}_{n}, & \text{when } e = 1.00,
    \end{cases}
\end{equation*}

In line with the scalar case, we observe that the eigenvalues of the saddle-point value parameters are significantly lower when $e = 0.85$ compared to $e = 1.0$. This indicates lower incentives for a takeover  when the system is open-loop stable $e < 1$ as opposed to unstable condition of $e \geq 1$. Notably, the parameter $\overline{P}_{k}^{0}$ consistently achieves a steady-state for both values of $e$, suggesting that the system will remain stable under the defender's control, regardless of the open-loop stability or instability of the system.
 
For the $n-$dimensional case, the takeover policy depends on the state $x$.
We simulate the system over 100 iterations with the initial state \(x_0 = \begin{bmatrix} 1 & 0 \end{bmatrix}^\top\) and present the average takeover policies in Figures~\ref{fig:FDC_pol_leq_ndim} and~\ref{fig:FDC_pol_geq_ndim}. 
In the mixed NE takeover (M-NE) scenario, for both \(e = 0.85\) and \(e = 1.0\) and \(\alpha = 0\) (defender in control), the probability of takeover increases for the defender and decreases for the adversary backward in time, indicating that the defender retains control while the adversary remains idle.
In scenarios alternating between pure and mixed NE, the players switch between these strategies throughout the horizon for both \(e = 0.85\) and \(e = 1.0\) with \(\alpha = 0\).

This numerical example illustrates the utility of the approximate saddle-point value parameters in determining the takeover strategies for each player. Moreover, it offers valuable insight into the system's behavior under specified costs and its stability properties.

\section{Conclusion and Future Directions}\label{sec:Conclusion}
In this work, we introduced \texttt{FlipDyn-Con}, a finite-horizon, zero-sum game of resource takeovers in discrete-time dynamical systems. Our key contributions include: deriving analytical expressions for saddle-point values and NE takeover strategies (pure and mixed) for general systems with known control policies; developing optimal linear state-feedback control policies for linear systems with quadratic costs and sufficient conditions for saddle-point existence; obtaining exact saddle-point values and NE strategies for scalar systems; and establishing bounds for saddle-point parameters and NE strategies for higher-dimensional systems. The practical relevance of our framework was demonstrated through a numerical study of a linear system under adversarial control.

Our future work will focus on expanding the \texttt{FlipDyn-Con} framework by incorporating partial state observability, and introducing bounded process and measurement noise to study its impact on the game.
Additionally, we plan to design a learning-based approach for the $n-$dimensional case and compare it with our approximate solution across various objectives and cost functions, enabling robustness and applicability of complex real-world systems.

\bibliographystyle{IEEEtran}
\bibliography{ref_TAC}


\appendices
\section{Proof of Theorem 1}\label{app:Proof_of_Theorem_1}
\begin{proof}
    We derive the NE takeover strategies and saddle-point value in the space of pure policies for $\alpha_k = 0$.
    The NE takeover strategies in the space of mixed policies directly follow from our prior work~\cite{FlipDyn_banik2022}.
    We omit the derivations for $\alpha_k = 1$ as they are analogous to the case of $\alpha_k = 0$. The $2 \times 2$ matrix game in~\eqref{eq:Cost_to_go_al0} gives rise to three possible cases of NE.
    
    i) \underline{Pure strategy}:
    
    Both the defender and adversary choose to remain idle (not takeover). 
    \noindent We begin by establishing the conditions under which the defender always chooses to play idle. 
    Under Assumption~\ref{ast:general_costs}, we compare the entries of $\Xi_{k+1}^{0}$ when the adversary also remains idle, which yields the condition:
    \begin{equation}
        \begin{aligned}
            v_{k+1}^{0} \leq v_{k+1}^{0} + d_{k}(x).
        \end{aligned}
    \end{equation}
    Similarly, the condition when adversary opts to takeover while the defender remains idle is given by:
    \begin{equation*}
        \begin{aligned}
            v_{k+1}^{1} & \leq v_{k+1}^{0} +  d_{k}(x), 
        \end{aligned}
    \end{equation*}
    \begin{equation*}
        \begin{aligned}
         \Rightarrow & v_{k+1}^{1} - v_{k+1}^{0} \leq d_k(x)
        \end{aligned}
    \end{equation*}
    Next, we determine the conditions under which the adversary always remain idle. Under Assumption~\ref{ast:general_costs}, when the defender chooses to takeover, we compare the entries of $\Xi_{k+1}^{0}$ and obtain: 
    \begin{equation*}
        \begin{aligned}
                v_{k+1}^{0} + d_{k}(x) \geq & 
                 v_{k+1}^{0}  + d_{k}(x) - a_{k}(x) \\ 
              \Rightarrow 0 \geq & -a_{k}(x),
        \end{aligned}
    \end{equation*}
    which always holds since $a_k(x) \geq 0$.
    Finally, when the adversary remains idle, the defender also remains idle if:
    \begin{equation*}
        \begin{aligned}
           v_{k+1}^{1} \leq v_{k+1}^{0} + a_{k}(x).
        \end{aligned}
    \end{equation*}
    The saddle-point value corresponding to the pure strategy in which both players remain idle is given by the entry $\Xi_{k+1}^{0}(1,1)$, which yields:
    \begin{equation*}
        V_{k}^{0}(x,u_k,\Xi_{k+1}^{0}) = g_k(x,0) + v_{k+1}^{0} + m_k(u_k).
    \end{equation*}
    
    ii) \underline{Pure strategy}:
    The defender chooses to remain idle, while the adversary chooses to takeover.
    \noindent We now derive the conditions under which the adversary opts to takeover. When the defender remains idle, the adversary prefers to take over if the following condition holds:
    \begin{equation*}
        \begin{aligned}
            v_{k+1}^{1} \geq v_{k+1}^{0} + a_{k}(x),
        \end{aligned}
    \end{equation*}
    If this inequality is satisfied, the adversary always chooses to takeover.
    The corresponding saddle-point value for this pure strategy is given by the entry $\Xi_{k+1}^{0}(1,2)$, which yields:
    \begin{equation*}
        V_{k}^{0}(x,u_k,\Xi_{k+1}^{0}) = g_k(x,0) + m_k(u_k) + v_{k+1}^{1} - a_k(x).
    \end{equation*}


    By collecting the saddle-point values of the game corresponding to both pure and mixed strategy~\cite{FlipDyn_banik2022} NE, we obtain the saddle-point value update equation over the finite-horizon $L$ in~\eqref{eq:Val_gen_al0}.
    Note that $g_k(x,0)$ and $m_k(u_k)$ represent the instantaneous state and control-dependent costs, respectively, and are not part of the zero-sum matrix in~\eqref{eq:V_k^0_cost_to_go}.
    The boundary conditions~\eqref{eq:b_cond_NE_Val_gen_FDC} imply that the saddle-point values at $k = L+1$ satisfy:
    \begin{equation*}
        \begin{aligned}
            V_{L+1}^{0}(x,\mathbf{0}_{m},\mathbf{0}_{2 \times 2}) & = g_{L+1}^{0}(x,0), \\ V_{L+1}^{1}(x,\mathbf{0}_{p},\mathbf{0}_{2 \times 2}) & = g_{L+1}^{1}(x,1).
        \end{aligned}
    \end{equation*}
\end{proof}

\section{Proof of Theorem 2}\label{app:Theorem_2}
\begin{proof}
    Under Assumptions~\ref{ast:linear_control_space} and~\ref{ast:takeover_costs}, if the adversary control policy $w_k^{*}(x)$ is known, the defender's control problem reduces to:
    \begin{equation}\label{eq:def_control_problem}
        \min_{K_k} v_{k+1}^{0} + x_k^{\tp}K_kM_kK_kx - \dfrac{x^{\tp}d_k\mathbb{I}_{n}xx^{\tp}a_k\mathbb{I}_{n}x}{v_{k+1}^{1*} - v_{k+1}^{0}},
    \end{equation}
    where $v_{k+1}^{1*} := x^{\tp}(E_k + H_kW_k^{*})^{\tp}P_{k+1}^{1}(E_k + H_kW_k^{*})x$ and $v_{k+1}^{0}$ is defined in~\eqref{eq:V_k_0}.
    Similarly, the adversary's control problem for a known defender policy $u_k^{*}(x)$ is given by:
    \begin{equation}\label{eq:adv_control_problem}
        \max_{W_k} v_{k+1}^{1} - x_k^{\tp}W_kN_kW_kx + \dfrac{x^{\tp}d_k\mathbb{I}_{n}xx^{\tp}a_k\mathbb{I}_{n}x}{v_{k+1}^{1} - v_{k+1}^{0*}},
    \end{equation}
    where $v_{k+1}^{0*} := x^{\tp}(E_k + B_kK_k^{*})^{\tp}P_{k+1}^{0}(E_k + B_kK_k^{*})x$, and $v_{k+1}^{1}$ is defined in~\eqref{eq:V_k_1}.
    
    Taking the first derivative of~\eqref{eq:def_control_problem} and~\eqref{eq:adv_control_problem} with respect to the player control gains $K_k$ and $W_k$, respectively, and solving the first-order optimality conditions, yields:
    \begin{equation}\label{eq:FOC_def}
        \begin{aligned}
            B^{\tp}_{k}P_{k+1}^{0}(E_{k} + B_{k}K_{k}) & + M_{k}K_{k} - \\ &\tfrac{a_{k}d_{k}(x^{\tp}x)^{2}B^{\tp}_{k}P_{k+1}^{0}(E_{k} + B_{k}K_{k})}{\left(v_{k+1}^{1*} - v_{k+1}^{0}\right)^{2}} = \mathbf{0}_{m \times n},
        \end{aligned}
    \end{equation}
    \begin{equation}\label{eq:FOC_adv}
        \begin{aligned}
            H^{\tp}_{k}P_{k+1}^{1}(E_{k} + H_{k}W_{k}) & - N_{k}W_{k} -  \\ & \tfrac{a_{k}d_{k}(x^{\tp}x)^{2}H^{\tp}_{k}P_{k+1}^{1}(E_{k} + H_{k}W_{k})}{(v_{k+1}^{1} - v_{k+1}^{0*})^{2}} = \mathbf{0}_{p \times n},
        \end{aligned}
    \end{equation}
    where $\mathbf{0}_{i \times j} \in \mathbb{R}^{i \times j}$ is a matrix of zeros. The terms 
    \begin{equation*}
        \tfrac{a_{k}d_{k}(x^{\tp}x)^{2}B^{\tp}_{k}P_{k+1}^{0}(E_{k} + B_{k}K_{k})}{\left(v_{k+1}^{1*} - v_{k+1}^{0}\right)^{2}} \text{ and } \tfrac{a_{k}d_{k}(x^{\tp}x)^{2}H^{\tp}_{k}P_{k+1}^{1}(E_{k} + H_{k}W_{k})}{(v_{k+1}^{1} - v_{k+1}^{0*})^{2}},
    \end{equation*}
    introduce non-linearity in $K_k$ and $W_k$ in~\eqref{eq:FOC_def} and~\eqref{eq:FOC_adv}, respectively. Such non-linearity inhibits the derivation of an optimal \emph{linear} control policy of the form~\eqref{eq:linear_sfdb_pol}. To address this, we introduce scalar parameters $\eta_{k,0} \in \mathbb{R}$ and $\eta_{k,1} \in \mathbb{R}$, satisfying:
    \begin{equation}\label{eq:eta_0_SC}
        \begin{aligned}
            &x^{\tp}\left((E_{k} + H_{k}W_{k}^{*})^{\tp}P_{k+1}^{1}(E_{k} + H_{k}W_{k}^{*}) \right. -  \\ & \left. 
            (E_{k} + B_{k}K_{k})^{\tp}P_{k+1}^{0}(E_{k} + B_{k}K_{k})\right)x = x^{\tp}x\dfrac{\sqrt{a_kd_k}}{\eta_{k,0}},
        \end{aligned}
    \end{equation}
    \begin{equation}\label{eq:eta_1_SC}
        \begin{aligned}
            &x^{\tp}\left((E_{k} + H_{k}W_{k})^{\tp}P_{k+1}^{1}(E_{k} + H_{k}W_{k}) \right. -  \\ & \left. 
            (E_{k} + B_{k}K_{k}^{*})^{\tp}P_{k+1}^{0}(E_{k} + B_{k}K_{k}^{*})\right)x = x^{\tp}x\dfrac{\sqrt{a_kd_k}}{\eta_{k,1}}.
        \end{aligned}
    \end{equation}
    Substituting~\eqref{eq:eta_0_SC} and~\eqref{eq:eta_1_SC} in~\eqref{eq:FOC_def} and~\eqref{eq:FOC_adv}, respectively, and solving for the parameterized control gains we obtain:
    \begin{equation}\label{eq:def_K_eta_0}
        K_k^{*} = -((1 - \eta_{k,0}^{2})B_{k}^{\tp}P_{k+1}^{0}B_{k} + M_k)^{-1}((1 - \eta_{k,0}^{2})B_{k}^{\tp}P_{k+1}^{0}E_k),
    \end{equation}
    \begin{equation}\label{eq:adv_W_eta_1}
        W_k^{*} = -((1 - \eta_{k,1}^{2})H_{k}^{\tp}P_{k+1}^{1}H_{k} - N_k)^{-1}((1 - \eta_{k,1}^{2})H_{k}^{\tp}P_{k+1}^{1}E_k).
    \end{equation}
    Substituting~\eqref{eq:def_K_eta_0} and~\eqref{eq:adv_W_eta_1} in~\eqref{eq:eta_0_SC} and~\eqref{eq:eta_1_SC}, respectively, yields an identical equation. This observation implies that, if a common parameter $\eta_{k}$ exists such that $\eta_{k} = \eta_{k,0} = \eta_{k,1}$, then the control policy pair~\eqref{eq:linear_stfb_def_mNE} and~\eqref{eq:linear_stfb_adv_mNE} satisfies the condition~\eqref{eq:suff_cond_eta}. The control policy pair $\{K_{k}^{*},W_{k}^{*}\}$ constitutes a mixed strategy NE takeover with the saddle-point values $V_{k}^{0}(x)$ and $V_{k}^{1}(x)$, provided they satisfy the conditions:
    \begin{equation*}
        \widetilde{P}_{k+1}(x) > d_kx^{\tp}x, \ \widetilde{P}_{k+1}(x) > a_kx^{\tp}x.
    \end{equation*}
    Substituting the dynamics~\eqref{eq:linear_dynamics} and the parameterized optimal control policies $(u_{k}^{*}(x), w_{k}^{*}(x))$ in~\eqref{eq:gen_mNE_eqns} and factoring out the state $x$, we obtain the conditions~\eqref{eq:LQ_mNE_cond1} and~\eqref{eq:LQ_mNE_cond2}. 

    Furthermore, substituting~\eqref{eq:eta_0_SC} and~\eqref{eq:eta_1_SC} in~\eqref{eq:FOC_def} and~\eqref{eq:FOC_adv}, respectively, taking the second derivative with respect to $K_{k}^{*}$ and $W_{k}^{*}$ and solving for the second-order conditions, we conclude that the controls are optimal provided:
    \begin{equation}\label{eq:SOO_eta_gen}
        (1 - \eta_{k}^{2})B_{k}^{\tp}P_{k+1}^{0}B_{k} + M_k \succ 0, \quad (1 - \eta_{k}^{2})H_{k}^{\tp}P_{k+1}^{1}H_{k} - N_k \prec 0.
    \end{equation}
    Given the quadratic costs~\eqref{eq:ST_MV_cost_Q}, as $\eta_{k} \to 0$, the second-order optimality condition~\eqref{eq:SOO_eta_gen} is always satisfied. Setting $\eta_{k} = 1$ in~\eqref{eq:SOO_eta_gen}, yields the limiting conditions. The obtained conditions verify/certify strong convexity in the control gain $K_k$ and strong concavity in $W_k$, ensuring the existence of a unique saddle-point equilibrium.  
\end{proof}

\section{Proof of Theorem 3}\label{app:Theorem_3}

\begin{proof}
    We will establish the proof only for the defender's control policy, as the derivation is analogous for the adversary's control policy. We start by examining the conditions in both~\eqref{eq:def_control_mnpNE} and~\eqref{eq:adv_control_mnpNE}, specifically:
    \[
        \widetilde{P}_{k+1}^{*}(x) > x^{\tp}d_k\mathbb{I}_{n}x, \text{ and } \widetilde{P}_{k+1}^{*}(x) > x^{\tp}a_k\mathbb{I}_{n}x.
    \]
    Under these conditions, along with those specified in~\eqref{eq:LQ_mNE_cond1}, \eqref{eq:LQ_mNE_cond2} and \eqref{eq:suff_cond_eta},   Theorem~\ref{th:linear_control_policy} yields mixed strategy NE takeover policies. To complete the remaining part of this proof, we proceed to derive the control policies for NE takeovers in pure strategies.

    i) \underline{Pure strategy}:
    The defender chooses to stay idle whereas the adversary chooses to takeover. 
    \noindent This takeover strategy is defined by the following conditions:
    \[\widetilde{P}_{k+1}^{*}(x) \leq x^{\tp}d_kIx, \text{ and } \widetilde{P}_{k+1}^{*}(x) > x^{\tp}a_k\mathbb{I}_{n}x.\]
    If the optimal adversary control policy $w_k^{*}(x)$ for the corresponding pure strategy NE takeover is known, the defender's control problem simplifies to:
    \begin{equation}\label{eq:def_control_problem_pNE}
        \min_{K_k} v_{k+1}^{1*} + x_k^{\tp}K_{k}^{\tp}M_{k}K_{k}x -x_k^{\tp}a_k\mathbb{I}_{n}x.
    \end{equation}
    Taking the first derivative of~\eqref{eq:def_control_problem_pNE} with respect to $K_k$, and subsequently applying the first-order optimality condition under the assumption $M_{k} \in \mathbb{S}_{+}^{m \times m}$, we obtain:
    \begin{equation*}
        \begin{aligned}
            & M_{k}K_{k}xx^{\tp} = \mathbf{0}_{m \times n}, \Rightarrow M_{k}^{-1}M_{k}K_{k}xx^{\tp} = \mathbf{0}_{m \times n}, \ \forall x \in \mathbb{R}^n \\
            & \Rightarrow K_k = \mathbf{0}_{m} = K_k^{*}(\eta_k = 1).
        \end{aligned}
    \end{equation*}
    This implies that the defender refrains from applying any control input due to a deterministic adversarial takeover at $k+1$. Notably, this condition of zero control gain aligns with setting $\eta_k = 1$ in~\eqref{eq:def_K_eta_0}.  

    ii) \underline{Pure strategy}:
    Both the defender and adversary opt to remain idle.   
    \noindent In this scenario, the takeover strategy is characterized by the following conditions: 
    \[
    \widetilde{P}_{k+1}^{*}(x) \leq x^{\tp}d_k\mathbb{I}_{n}x, \text{ and } \widetilde{P}_{k+1}^{*}(x) \leq x^{\tp}a_k\mathbb{I}_{n}x.
    \]
    Given the absence of an adversary control term in determining the saddle-point value of the game, the defender's control problem  simplifies to:
    \begin{equation}\label{eq:def_control_problem_pNE_idle}
        \min_{K_k} v_{k+1}^{0} + x_k^{\tp}K_{k}^{\tp}M_{k}K_{k}x.
    \end{equation}
    Taking the first derivative of~\eqref{eq:def_control_problem_pNE_idle} with respect to $K_k$, and solving for the first-order optimality condition, we obtain:
    \begin{equation*}
        \begin{aligned}
            & B_{k}^{\tp}P_{k+1}^{0}(E_k + B_{k}K_k) = -M_kK_k, \\
            & \Rightarrow K_k = (M_k + B^{\tp}P_{k+1}^{0}B_k)^{-1}B^{\tp}_kP_{k+1}^{0}E_k := K_k^{*}(\eta_k = 0).
        \end{aligned}
    \end{equation*}
    This control policy pertains to a single-player control problem, given that the \texttt{FlipDyn} state deterministically remains at $\alpha_{k+1} = 0$.  Furthermore, this control policy corresponds to setting $\eta_k = 0$ in~\eqref{eq:def_K_eta_0}. 
\end{proof}

\section{Proof of Proposition 1}\label{app:Proposition_1}

\begin{proof}
    A permissible parameter $\eta_{k}$  satisfying~\eqref{eq:suff_cond_eta} corresponds to a control policy pair $\{u_{k}^{*}(x),w_{k}^{*}(x)\}$ that constitutes a mixed strategy NE takeover with saddle-point values $V_{k}^{0}(x)$ and $V_{k}^{1}(x)$. For such a control policy pair and $\eta_{k}$, the following condition must hold:
    \begin{equation*}
        \widetilde{P}_{k+1}(x) > x^{\tp}d_k\mathbb{I}_{n}x, \ \widetilde{P}_{k+1}(x) > x^{\tp}a_k\mathbb{I}_{n}x.
    \end{equation*}
    Since a lower bound on $\widetilde{P}_{k+1}(x)$ is equivalent to the condition~\eqref{eq:suff_cond_eta}, we substitute the right-hand side of~\eqref{eq:suff_cond_eta} into the prior stated conditions to obtain:
    \begin{equation*}
        x^{\tp}x\dfrac{\sqrt{a_k d_k}}{\eta_k}x > d_{k}x^{\tp}x, \ x^{\tp}x\dfrac{\sqrt{a_k d_k}}{\eta_k} > a_{k}x^{\tp}x.
    \end{equation*}
    By eliminating the state $x$ and combining the terms, we arrive at~\eqref{eq:eta_bound}.
\end{proof}

\section{Proof of Corollary 1}\label{app:Corollary_1}
\begin{proof}
We begin the proof by determining the NE takeover in both pure and mixed strategies, and computing the corresponding saddle-point value parameter for the \texttt{FlipDyn} state of $\alpha = 0$. We substitute the quadratic costs~\eqref{eq:cost_form_scalar}, linear dynamics~\eqref{eq:linear_dynamics_LOC_DA}, and the obtained optimal control policies~\eqref{eq:def_control_mnpNE} and~\eqref{eq:adv_control_mnpNE} in the term $\widetilde{P}_{k+1}(x)$ from~\eqref{eq:gen_mNE_eqns} to obtain:
\begin{equation*}
    \begin{aligned}
        \tilde{P}_{k+1}(x) & := \left((E_x + H_kW_k^{*}(\eta_k))^{2}\mathbf{p}_{k+1}^{1} - \right. \\ & \hphantom{= \mathbf{p}_{k+1}^{1} } \left. (E_x + B_kK_k^{*}(\eta_k))^{2}\mathbf{p}_{k+1}^{0}\right)x^{2}, \\
         & = \left(\mathbf{p}_{k+1}^{1} \dfrac{N^2_{k} - \hat{\eta}_{k}H_{k}^{2}\mathbf{p}_{k+1}^{1} + \hat{\eta}_{k}H_{k}^{2}\mathbf{p}_{k+1}^{1}}{(N_{k} - \hat{\eta}_{k}H^{2}_{k}\mathbf{p}_{k+1}^{1})^{2}} - \right. \\ & \left. \hphantom{ = 000} \mathbf{p}_{k+1}^{0}\dfrac{M_{k} + \hat{\eta}_{k}B_{k}^{2}\mathbf{p}_{k+1}^{0} - \hat{\eta}_{k}B_{k}^{2}\mathbf{p}_{k+1}^{0}}{(M_{k} + \hat{\eta}_{k}B^{2}_{k}\mathbf{p}_{k+1}^{0})^{2}}\right)E^2_{k}x^2, \\
        & = \check{\mathbf{p}}_{k+1}x^{2}
    \end{aligned}
\end{equation*}
Substituting $\check{\mathbf{p}}_{k+1}$ and takeover costs~\eqref{eq:takeover_costs} in~\eqref{eq:def_TP_gen_FDC_al0} and~\eqref{eq:adv_TP_gen_FDC_al0}, we obtain the NE takeover strategies presented in~\eqref{eq:def_TP_1D_FDC_al0} and~\eqref{eq:adv_TP_1D_FDC_al0}, respectively. Notably, as observed in Theorem~\ref{th:NE_Val_gen_FDC}, the NE takeover strategies for the \texttt{FlipDyn} state of $\alpha_k = 1$ can be also be obtained by taking the complementary of~\eqref{eq:def_TP_1D_FDC_al0} and~\eqref{eq:adv_TP_1D_FDC_al0}, resulting in~\eqref{eq:def_TP_1D_FDC_al1} and~\eqref{eq:adv_TP_1D_FDC_al1}, respectively. 

To obtain a recurrence relation for the parameter $\mathbf{p}_{k}^{0}$, we substitute the linear dynamics~\eqref{eq:linear_dynamics_LOC_DA} along with quadratic costs~\eqref{eq:cost_form_scalar}, takeover costs~\eqref{eq:takeover_costs}. This yields
\begin{align*}\label{eq:Val_1D_al0}
    \mathbf{p}_{k}^{0}x^{2} = 
    \begin{cases}
        \begin{aligned}
            & (G^{0}_k + d_k)x^2 -\frac{d_{k}a_{k}x^4}{\check{\mathbf{p}}_{k+1}x^2} \\[2pt] & + (K_{k}^{*}(\eta_k)^{2}M_{k} + \check{B}_k(\eta_{k})^{2}\mathbf{p}^{0}_{k+1})x^2,
        \end{aligned} \hfill \text{if } \begin{matrix}
                {\check{\mathbf{p}}_{k+1}} > d_k, \\
                {\check{\mathbf{p}}_{k+1}} > a_k,
            \end{matrix} \\[10pt] 
        \begin{aligned}
            & (G^{0}_k+ \check{W}_k(\eta_{k})^{2}\mathbf{p}^{1}_{k+1} - a_{k})x^2,
        \end{aligned} \hfill \text{if } \begin{matrix}
                {\check{\mathbf{p}}_{k+1}} \leq d_k, \\
                {\check{\mathbf{p}}_{k+1}} > a_k,
            \end{matrix} \\[10pt] 
         \begin{aligned}
            & (G^{0}_k + K_{k}^{*}(0)^2M_{k} + \check{B}_k(0)^{2}\mathbf{p}^{0}_{k+1})x^2,
        \end{aligned}
         \hfill \text{otherwise}.
    \end{cases} 
\end{align*}
Substituting the control gains $K_{k}^{*}(\eta_k)$~\eqref{eq:def_control_mnpNE} and $W_{k}^{*}(\eta_k)$~\eqref{eq:adv_control_mnpNE} and factoring out the term $x^2$, we arrive at~\eqref{eq:Val_1D_al0}. Employing analogous substitutions for the \texttt{FlipDyn} state of $\alpha_k = 1$, we obtain~\eqref{eq:Val_1D_al1}.

Condition~\eqref{eq:suf_cond_1d} corresponds to a second-order optimality condition for the policy pair $\{u_{k}^{*}(x),w_{k}^{*}(x)\}$  derived for a scalar dynamical system. This condition ensures that the control policies form a saddle-point equilibrium.
\end{proof}

\section{Proof of Corollary 2}\label{app:Corollary_2}

\begin{proof}
    We begin the proof by determining the NE takeover in pure and mixed strategies of the \texttt{FlipDyn} state of $\alpha = 0$. We substitute the takeover cost~\eqref{eq:takeover_costs} and the terms  from~\eqref{eq:exact_sol_non_in_x} in~\eqref{eq:def_TP_gen_FDC_al0} and~\eqref{eq:adv_TP_gen_FDC_al0}, to obtain the NE takeover policies in~\eqref{eq:def_TP_nD_FDC_al0} and~\eqref{eq:adv_TP_nD_FDC_al0}, respectively. Analogous to the scalar case, the NE takeover strategies in~\eqref{eq:def_TP_nD_FDC_al1} and~\eqref{eq:adv_TP_nD_FDC_al1} for the \texttt{FlipDyn} state of $\alpha = 1$ are the complementary takeover strategies of the \texttt{FlipDyn} state $\alpha = 0$. 

    To determine the saddle-point value parameters for the \texttt{FlipDyn} state of $\alpha = 0$, we substitute~\eqref{eq:exact_sol_non_in_x}, discrete-time linear dynamics~\eqref{eq:linear_dynamics_LOC_DA}, quadratic costs~\eqref{eq:ST_MV_cost_Q} and takeover costs~\eqref{eq:takeover_costs} in~\eqref{eq:Val_gen_al0} and factor out the state $x$ to obtain~\eqref{eq:Val_nD_al0}. Through similar substitutions and factorization we can obtain~\eqref{eq:Val_nD_al1} corresponding to the \texttt{FlipDyn} state of $\alpha = 1$.
\end{proof}

\section{Proof of Lemma 1}\label{app:Lemma_1}

\begin{proof}
    From~\eqref{eq:linear_stfb_def_mNE}, a linear defender control policy gain parameterized by a scalar $\underline{\eta}_{k}$, is given by:
    \begin{equation}
        \label{eq:linear_stfb_def_mNE_ub}
        \begin{aligned}
             K_k^{*}(\underline{\eta}_{k}) = -(\vartheta(\underline{\eta}_{k})B_{k}^{\tp}P_{k+1}^{0}B_{k} + M_k)^{-1}(\vartheta(\underline{\eta}_{k})B_{k}^{\tp}P_{k+1}^{0}E_k),
        \end{aligned}
    \end{equation}
    where $\vartheta(c) := 1 - c^{2}$. Likewise, from~\eqref{eq:linear_stfb_adv_mNE}, a linear adversary control policy gain parameterized by a scalar $\overline{\eta}_{k}$, is given by:
    \begin{equation}
        \label{eq:linear_stfb_adv_mNE_lb}
        \begin{aligned}
        W_k^{*}(\overline{\eta}_{k}) = -(\vartheta(\overline{\eta}_{k})H_{k}^{\tp}P_{k+1}^{1}H_{k} - N_k)^{-1}(\vartheta(\overline{\eta}_{k})H_{k}^{\tp}P_{k+1}^{1}E_k).
        \end{aligned}
    \end{equation}
    Upon substituting the condition~\eqref{eq:suff_cond_eta_bounds} in~\eqref{eq:def_control_problem} and~\eqref{eq:adv_control_problem} and solving for the second-order optimality condition (similar to Theorem~\ref{th:linear_control_policy}) yields~\eqref{eq:gen_SC_KW_bounds}, which certifies a saddle-point 
    equilibrium. 

    Recall that any control policy pair $\{K_{k},W_{k}\}$ that constitutes a mixed strategy NE takeover to both the saddle-point values $V_{k}^{0}(x)$ and $V_{k}^{1}(x)$ must satisfy the conditions:
    \begin{equation*}
        \widetilde{P}_{k+1}(x) > d_kx^{\tp}x, \ \widetilde{P}_{k+1}(x) > a_kx^{\tp}x.
    \end{equation*}
    Thus, upon substituting the linear dynamics~\eqref{eq:linear_dynamics} and the optimal control gains $\{K_{k}^{*}(\underline{\eta}_{k}), W_{k}^{*}(\overline{\eta}_{k})\}$ in~\eqref{eq:gen_mNE_eqns} and factoring out the state $x$, we obtain the conditions~\eqref{eq:LQ_mNE_cond1_bounds} and~\eqref{eq:LQ_mNE_cond2_bounds}. 
    
    Next, we will only establish~\eqref{eq:P_0_ub}, as the derivation for~\eqref{eq:P_1_lb} is analogous. Under a mixed strategy NE takeover, we substitute the quadratic costs~\eqref{eq:ST_MV_cost_Q}, discrete-time linear dynamics~\eqref{eq:linear_dynamics_LOC_DA} and the defender control~\eqref{eq:linear_stfb_def_mNE_ub} in~\eqref{eq:Val_gen_al0} to obtain:
    \begin{equation*}
         \begin{aligned} x^{\tp}P_{k}^{0}x =
            &x^{\tp}\left(G^{0}_k + d_{k}\mathbb{I}_{n} + K_{k}^{*}(\underline{\eta}_k)^{\tp}M_kK_{k}^{*}(\underline{\eta}_k) 
  \right)x +  \\
            & x^{\tp}\left( \check{B}_{k}(\underline{\eta}_k)^{\tp}P^{0}_{k+1}\check{B}_{k}(\underline{\eta}_k) \right)x -  \\ & \hspace{-15pt}\dfrac{x^{\tp}a_{k}\mathbb{I}_{n}xx^{\tp}d_{k}\mathbb{I}_{n}x}{x^{\tp}\underbrace{\left(\check{W}_{k}(\overline{\eta}_k)^{\tp}P^{1}_{k+1}\check{W}_{k}(\overline{\eta}_k) - \check{B}_{k}(\underline{\eta}_k)^{\tp}P^{0}_{k+1}\check{B}_{k}(\underline{\eta}_k)\right)}_{\mathbf{P}_{k+1}}x}.
        \end{aligned}
    \end{equation*}
    Using condition~\eqref{eq:suff_cond_eta_bounds}, we bound the term containing $\mathbf{P}_{k+1}$ by
    \begin{equation*}
        \begin{aligned}
            \dfrac{x^{\tp}a_{k}\mathbb{I}_{n}xx^{\tp}d_{k}\mathbb{I}_{n}x}{x^{\tp}\mathbf{P}_{k+1}x} & \leq \underline{\eta}_{k}\dfrac{x^{\tp}a_{k}\mathbb{I}_{n}xx^{\tp}d_{k}\mathbb{I}_{n}x}{x^{\tp}x \sqrt{a_{k}d_{k}}},\\
            & \leq \underline{\eta}_{k}x^{\tp}x \sqrt{a_{k}d_{k}}.
        \end{aligned}
    \end{equation*}
    Substituting this bound in $ x^{\tp}P_{k}^{0}x$ and factoring out the state $x$, we obtain~\eqref{eq:P_0_ub}. 

\end{proof}

\section{Proof of Corollary}\label{app:Corollary_3}

\begin{proof}{[Outline]}
    Similar to the proofs in the prior sections, we begin the proof by determining the NE takeover in pure and mixed strategies for the \texttt{FlipDyn} state of $\alpha = 0$. 
    We substitute the quadratic costs~\eqref{eq:ST_MV_cost_Q}, linear dynamics~\eqref{eq:linear_dynamics_LOC_DA}, and linear control gains~\eqref{eq:linear_stfb_def_mNE_ub} and~\eqref{eq:linear_stfb_adv_mNE_lb} in the term $\widetilde{P}_{k+1}(x)$ with the approximate saddle-point value parameters $\overline{P}_{k+1}^{0}$ and $\overline{P}_{k+1}^{1}$ from~\eqref{eq:gen_mNE_eqns} to obtain:
    \begin{equation*}
        \begin{aligned}
            \tilde{P}_{k+1}(x) & := \overline{V}_{k+1}^{1}(\check{W}_{k}(\overline{\eta}_{k})x) - \overline{V}_{k+1}^{0}(\check{B}_{k}(\underline{\eta}_{k})x), \\
             & = x^{\tp}\left(\check{W}_{k}^{\tp}(\overline{\eta}_{k})\overline{P}_{k+1}^{1}\check{W}_{k}(\overline{\eta}_{k}) \right. \\
            & \hphantom{= \mathbf{p}_{k+1}^{1} }  \left. - \check{B}_{k}(\underline{\eta}_{k})^{\tp}\overline{P}_{k+1}^{0}\check{B}_{k}(\underline{\eta}_{k}) \right)x, \\
            & = x^{\tp}\overline{\mathbf{P}}_{k+1}x.
        \end{aligned}
    \end{equation*}
    Substituting the takeover cost~\eqref{eq:takeover_costs} and $x^{\tp}\overline{\mathbf{P}}_{k+1}x$ in~\eqref{eq:def_TP_gen_FDC_al0} and~\eqref{eq:adv_TP_gen_FDC_al0}, we obtain the NE takeover policies in~\eqref{eq:def_TP_nD_FDC_al0_bounds} and~\eqref{eq:adv_TP_nD_FDC_al0_bounds}, respectively. The approximate NE takeover strategies of the \texttt{FlipDyn} state $\alpha = 1$ are complementary to $\alpha = 0$, presented in~\eqref{eq:def_TP_nD_FDC_al1_bounds} and~\eqref{eq:adv_TP_nD_FDC_al1_bounds}.

    To determine the approximate saddle-point value parameters under a mixed strategy NE takeover of the \texttt{FlipDyn} state of $\alpha = 0$, we substitute the upper bound~\eqref{eq:P_0_ub} from Lemma~\ref{lem:linear_control_policy_bounds} and replace $P_{k+1}^{0}$ with $\overline{{P}}_{k+1}^{0}$. Under a pure strategy NE takeover, we substitute the quadratic costs~\eqref{eq:ST_MV_cost_Q}, discrete-time linear dynamics~\eqref{eq:linear_dynamics_LOC_DA} and the adversary linear state-feedback control ~\eqref{eq:linear_stfb_adv_mNE_lb} to obtain the approximate saddle-point value parameters. Combining both the solutions from the mixed and pure strategy NE takeover, we obtain~\eqref{eq:Val_nD_al0_bounds}. 
\end{proof}

\end{document}